\documentclass[10pt,journal,compsoc]{IEEEtran}
\usepackage{graphicx}
\usepackage{algorithm}
\usepackage{algpseudocode}
\usepackage{subfigure}
\usepackage{epstopdf}
\usepackage{booktabs}
\usepackage{amsmath}
\usepackage[OT1]{fontenc} 
\ifCLASSOPTIONcompsoc
  \usepackage[nocompress]{cite}
\else
  \usepackage{cite}
\fi
\ifCLASSINFOpdf
\else
\fi
\usepackage{xcolor}
\usepackage[normalem]{ulem}

\begin{document}
%
\title{EmgAuth: Unlocking Smartphones with EMG Signals}
%
%
%
%

\author{Boyu Fan,
        Xiang Su,~\IEEEmembership{Member,~IEEE},
        Jianwei Niu,~\IEEEmembership{Senior Member,~IEEE}
        and Pan Hui,~\IEEEmembership{Fellow,~IEEE}
\IEEEcompsocitemizethanks{\IEEEcompsocthanksitem Boyu Fan is with the Department of Computer Science, University of Helsinki,
Helsinki 00100, Finland. E-mail: boyu.fan@helsinki.fi

\IEEEcompsocthanksitem Xiang Su is with the Department of Computer Science, University of Helsinki, Helsinki and Center of Ubiquitous Computing, University of Oulu, Finland. 
E-mail: xiang.su@helsinki.fi
\IEEEcompsocthanksitem Jianwei Niu is with the School of Computer Science and Engineering, Beihang University, Beijing, China.
E-mail: niujianwei@buaa.edu.cn
\IEEEcompsocthanksitem P. Hui is with the Department of Computer Science, University of Helsinki, Helsinki 00100, Finland, and also with the Department of Computer Science and Engineering, Hong Kong University of Science and Technology, Hong Kong. E-mail: panhui@cse.ust.hk
}
\thanks{Manuscript received Fed 15, 2021}
}

\IEEEtitleabstractindextext{%
\begin{abstract}
  Screen lock is a critical security feature for smartphones to prevent unauthorized access. Although various screen unlocking technologies, including fingerprint and facial recognition, have been widely adopted, they still have some limitations. For example, fingerprints can be stolen by special material stickers and facial recognition systems can be cheated by 3D-printed head models. In this paper, we propose EmgAuth, a novel electromyography(EMG)-based smartphone unlocking system based on the Siamese network. EmgAuth enables users to unlock their smartphones by leveraging the EMG data of the smartphone users collected from Myo armbands. When training the Siamese network, we design a special data augmentation technique to make the system resilient to the rotation of the armband, which makes EmgAuth free of calibration. We conduct extensive experiments including 53 participants and the evaluation results verify that EmgAuth can effectively authenticate users with an average true acceptance rate of 91.81\% while keeping the average false acceptance rate of 7.43\%. In addition, we also demonstrate that EmgAuth can work well for smartphones with different screen sizes and for different scenarios when users are placing smartphones at different locations and with different orientations. EmgAuth shows great promise to serve as a good supplement for existing screen unlocking systems to improve the safety of smartphones.
\end{abstract}

}

\maketitle

\IEEEdisplaynontitleabstractindextext

%
\IEEEpeerreviewmaketitle

\IEEEraisesectionheading{\section{Introduction}\label{sec:introduction}}

\IEEEPARstart{T}{he} safety of smartphones is very critical as devices store lots of individual private data, from emails to e-wallet payment details. Fortunately, screen lock helps us protect personal information from being accessed by others. A research shows that American people check their phones 96 times a day \cite{smartphoneusage}, which means that the smartphone is unlocked every ten minutes. However, it is not convenient to directly enter password when facing such a high frequency of use. Researchers began to develop more effective technologies to unlock smartphones. In particular, biometric-based technologies, including fingerprint, facial recognition, and iris\cite{daugman2009iris}, have gradually replaced the traditional password-based methods\cite{clarke2005authentication} in recent years, providing better convenience for users. Fingerprint and facial recognition are widely used in smartphones unlocking, however, there are still have certain security risks. For example, fingerprints can be easily obtained with packing tapes. Facial recognition can also be deceived, and studies have shown that 3D printed head models plus taped glasses can easily fool Apple's Face ID authentication system\cite{faceid}.  

Different from the above-mentioned biometric features, EMG signals, collected by placing electrodes on the skin to detect the electrical activity of muscles, show unique features for individuals and therefore has great potential for smartphone authentication\cite{sahoo2012multimodal}. In particular, we observe that when different people pick up their smartphones, the speed, wrist movement, fingers movement and the positions they grab smartphones are generally different from each other. In contrast, for a certain person, the movement when picking up a smartphone is generally consistent, which can be attributed to the memory of one's muscles accumulated over a long period of time. Compared with fingerprint and face-based authorisation methods, EMG-based methods are safer because EMG signals are dynamic and therefore harder to be obtained by others.

Based on these observations and analyses, we propose smartphone unlocking utilizing EMG signals. The uniqueness of the EMG signals and the consistency of motion movements are crucial for EMG-based authentication. Although there are some existing works that utilize EMG signals to unlock smartphones \cite{yamaba2015authentication}\cite{yamaba2017evaluation}, they require users to make a series of pre-defined gestures. In addition, the systems to collect EMG signals, such as the Myo armbands, need to be placed in the same position on the arm for both training and testing stages. These constraints limit the applicability of the EMG-based unlocking system as well as other EMG-based applications\cite{sathiyanarayanan2016myo}\cite{xu2016development}.

In this paper, we propose EmgAuth, a new EMG-based smartphone unlocking system based on the Siamese network. EmgAuth utilizes data collected from the Myo armband and allows users to unlock their smartphones when picking up and watching their smartphones without making any pre-defined gestures. A convolutional Siamese network is proposed to extract EMG features and achieve few shot learning. More importantly, when training the Siamese network, we design a special data augmentation technique to make the system resilient to the rotation of the armband. These two improvements significantly enhance the usability of the unlocking system. We implement EmgAuth on Android smartphones and recruit 53 participants to collect training and testing data for EmgAuth. 

We conduct a series of experiments to choose the proper parameters of EmgAuth, including the hyperparameters of the Siamese network and the threshold of the classifier. The results of cross validation demonstrate that EmgAuth can achieve good authentication accuracy in a real-time manner. It can authenticate users with an average true acceptance rate of 91.81\% while keeping an average false acceptance rate of 7.43\%, and the overall accuracy reaches 92.06\%. Experiments show that EmgAuth is rotation-independent. We also discuss some influencing factors in real-world scenarios to verify the feasibility of EmgAuth, such as the non-sitting scene, the impact of smartphone shape, the performance for left-handed users, etc. In addition, the authentication latency of EmgAuth that runs on an Android smartphone is about 0.16 s, which fulfills the requirement of real-time unlocking.

This paper is an extended version of \cite{fan2020emgauth} with a more detailed literature study, an upgraded Android-implemented prototype system, a bigger dataset and a deeper analysis to investigate the effect of rotation-independency and non-sitting scenarios.
The contributions of this paper are summarized as follows:

\begin{itemize}
  \item EmgAuth, a system that unlocks smartphones by natural motions based on EMG signals and Siamese network. 
  This is one of the first research efforts that combines EMG signals with deep learning to unlock smartphones.
  \item A novel method based on the structure of Myo armband to make EmgAuth resilient to armband rotation. With this method, users do not need to calibrate or remember the position of their armbands.
  \item Extensive experiments to verify the feasibility and reliability in different conditions.
\end{itemize}


The remainder of this paper is organized as follows. In Section 2, we discuss related work about common biometric authentication methods, EMG-based applications, and the Siamese network. Section 3 details our EmgAuth system architecture and each module. We then describe our dataset and provide the experimental results of our system in Section 4. In Section 5, we discuss some influencing factors in real-world scenarios. 
In Section 6, we discuss the advantages of EmgAuth with some practical issues and highlight our contributions. Finally, we conclude the paper with discussing limitations and future research directions in Section 7.

\section{Related work}
In this section, we discuss existing literature studies that relate to our work, including biometric authentication, EMG-based applications, and the Siamese network.
\subsection{Biometric Authentication}
Biometric authentication is widely used in daily life applications, such as transactions and user device login. Among various authentication methods, fingerprints are one of the most widely-used technologies. Anil Jain et al.\cite{jain1997line} first described the design and implementation of an online fingerprint authentication system. An alignment-based elastic matching algorithm is developed to find the correspondences between minutiae in the input image and the stored template. Facial recognition is another popular technology for identity authentication. Sun et al. \cite{sun2014deep} combined deep learning techniques with face identification. They used deep convolutional neural networks to learn features to reduce intra-personal variations while enlarging inter-personal differences and the accuracy can achieve a value of 99.15\%. Besides, iris and voice are also used in mobile device authentication \cite{2015Iris}\cite{2015Voice}.

In addition to physiological characteristics related methods, behavioral characteristics also attract attention from researchers. Keystroke dynamics is used as a kind of biometrics for authentication\cite{banerjee2012biometric}. Monrose et al.\cite{monrose2000keystroke} innovatively proposed a new authentication method based on analyzing habitual rhythm patterns when users type. They present data extraction methods, as well as classification strategies to achieve user authentication and the accuracy can reach 92.14\%. Electrocardiographic(ECG), the signals of the electrical activity of the heart, can also be used for authentication. Arteaga et al. \cite{arteaga2015ecg} first used ECG biometric signals to achieve authentication on mobile devices and the algorithm has 1.41\% false acceptance rate and 81.82\% true acceptance rate. Gait, hand-waving, signature and even the interaction with touchscreens are also used to enable authentication\cite{gafurov2006biometric}\cite{shrestha2013wave}\cite{sae2014online}\cite{frank2012touchalytics}.

\subsection{EMG-based applications}
EMG records the movement of muscles. Based on the simple fact that whenever a muscle contracts, a burst of electric activity is generated which propagates through adjacent tissue and bone which can then be recorded from neighboring skin areas. Therefore, EMG signals are widely used in medicine\cite{merletti2010advances}, control\cite{rechy2011stages}, human-computer interaction\cite{ahsan2009emg} and games\cite{zhang2009hand}. Kiguchi et al.\cite{kiguchi2012emg} used EMG signals to control an upper-limb power-assist exoskeleton robot, which is easy simple, human-like and adaptable to any user. EMG-based hand gesture identification can help develop a better human-computer interaction interface. In \cite{ahsan2011electromygraphy}, Ahsan described the process of detecting different hand gestures using an artificial neural network (ANN). They used a series of statistical methods to extract features and then feed these feature vectors to ANN to obtain a classification result. EMG signals are also combined with other sensors to achieve accurate control. Yoo et al.\cite{Yoo2014Effects} proposed an input device for a virtual reality game, which is based on EMG and accelerometers. The results show the device can offer good experiences for players. In addition, Myo armbands are one of the most popular devices for EMG-related research because of the portability and efficient data transmission mechanism\cite{becker2018touchsense}\cite{benalcazar2017hand}\cite{abreu2016evaluating}. Comparing with \cite{shin2017study}\cite{shioji2018personal}\cite{yamaba2019introduction}\cite{lu2020study}, EmgAuth does not need extra training datasets and is resilient to the positions of EMG sensors, and does not need users to conduct specific actions. Moreover, this is the first EMG-based system for smartphone unlocking with both hardware and software.

\subsection{The Siamese network}
Deep neural networks have excellent performance in the fields of image classification, speech recognition, and natural language processing. They can automatically extract features from large-scale data rather than conducting feature engineering manually. Many network structures are proposed to deal with different kinds of tasks\cite{xie2017lg}\cite{xu2017d}\cite{collobert2008unified}. In general deep learning methods, each category must have a very large amount of data to train a good model, which is not suitable for small datasets. Siamese network was first introduced by Bromley et al. to solve the problem of signature verification. They designed two identical sub-networks to extract features and combine them with a layer that computes the distance between the two outputs. Thus, it does not need a large dataset to learn, but just learns the difference between a pair. Inspired by them, many researchers leverage the Siamese network structure in various kinds of fields. Bertinetto et al.\cite{bertinetto2016fully} designed a novel fully-convolutional Siamese network trained end-to-end on the ILSVRC15 dataset for object detection in video. In \cite{zhang2016siamese}, the authors trained a Siamese network to enable human identification based on gait recognition. Siamese network is also the main technique in one-shot learning, Koch et al.\cite{koch2015siamese} used a Siamese neural network for one-shot image recognition, which does not need a very large dataset. Jianbo et al.\cite{2018One} leveraged convolutional Siamese neural network for fine-grained relation extraction, the result shows that this network can effectively extract features with limited samples.

\section{System Design and Components}
This section introduces the hardware and architecture of our EmgAuth system and the three main components, including data segmentation, the Siamese network, and the unlocking simulation system. We also detail the novel method to make EmgAuth resilient to the rotation of the Myo armband.

\subsection{System Architecture}
EmgAuth consists of components deployed on a Myo armband and an Android smartphone. Myo armband is a device collecting EMG signals. It has eight channels, corresponding to eight sensors in different positions. Each channel has a sample rate of 200 Hz and the data can be transferred over Bluetooth. Users' EMG signals can be easily retrieved by wearing the device on their arms. We design two Android applications. The first one is for collecting and labelling data, and the second is an unlocking simulation application. The deep learning model trained on a GPU server is then deployed to an Android smartphone and runs using TensorFlow Mobile.

\begin{figure}[]
    \setlength{\abovecaptionskip}{-0.cm}
    \setlength{\belowcaptionskip}{-0.cm}
    \centering
    \includegraphics[width=0.48\textwidth]{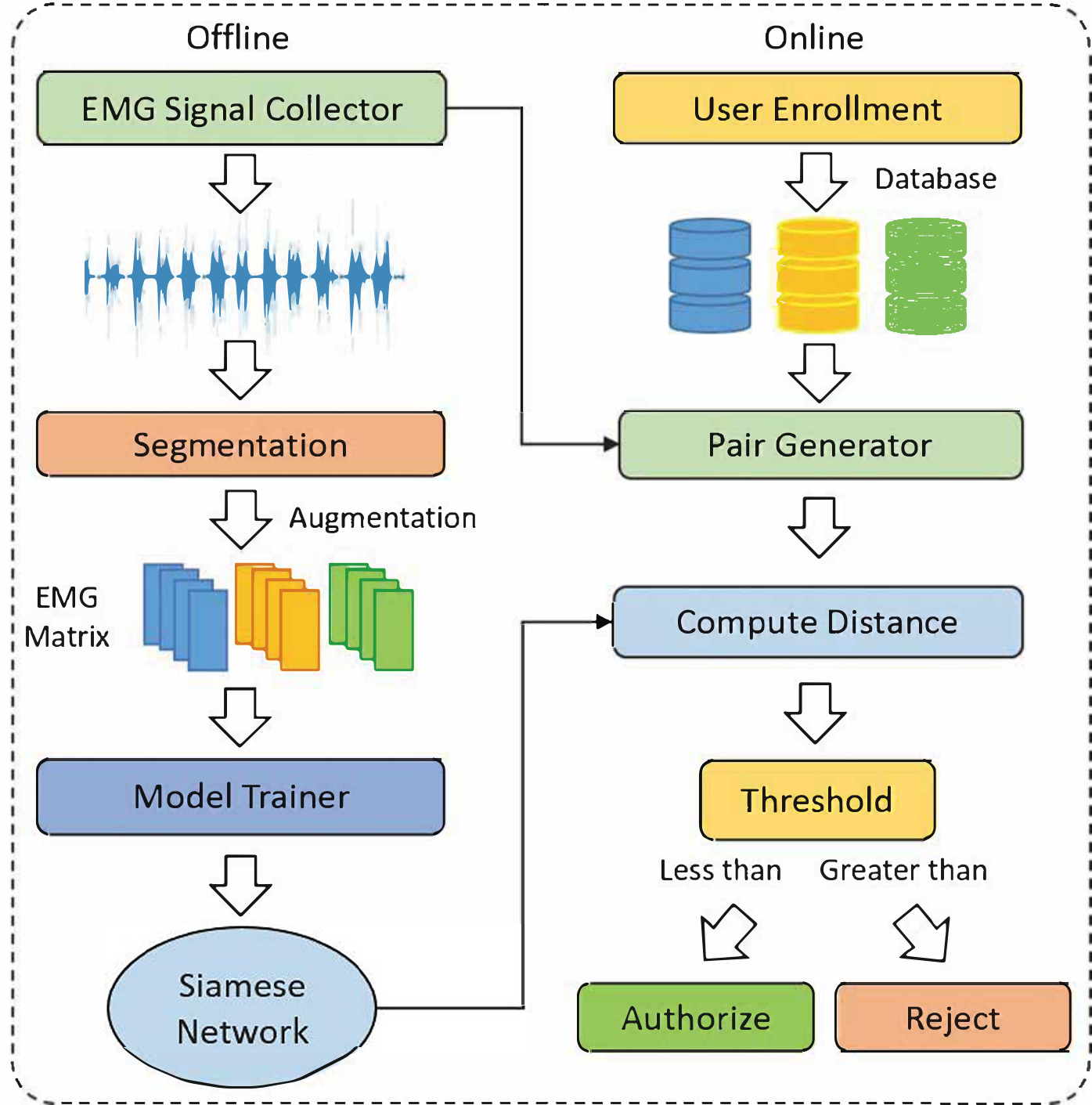}
    \caption{System architecture of EmgAuth.}
    \label{systemarchitecture}
\end{figure}

Figure \ref{systemarchitecture} presents the architecture of the system. The left side presents the offline model training. The data is collected by a Myo armband and transferred to the smartphone by Bluetooth in real-time. We label the different motions by clicking the corresponding buttons. After we get the labeled data, we conduct data segmentation to extract the valid EMG signals and make pairs to prepare training data. Data augmentation is also conducted to expand the dataset, and for achieving rotation-independence. Next, during the model training step, all pairs are fed into the neural network to train a convolutional Siamese neural network.

The right side of Figure \ref{systemarchitecture} describes the process of online authentication. We transplant the trained Siamese neural network to an Android smartphone so we can evaluate the performance in real scenarios. Similar to most authentication systems, the first step that the user needs to do is enrollment. The enrollment phase requires four sets of motions with the user only needing to execute each motion once. The system stores the EMG signals of these motions on the database and names them by the user's name and corresponding executed motion as identifiers. Next, when the user picks up the smartphone, the new EMG signal produced from the process will be compared with the previously stored signal and put into the model that we train in the offline phase. The Siamese neural network computes the distance of the input EMG pairs. If the output is less than the pre-defined threshold, the user will be successfully authorized and the smartphone unlocks, otherwise the user will be rejected. 

\begin{figure}
    \setlength{\abovecaptionskip}{-0.cm}
    \setlength{\belowcaptionskip}{-0.cm}
    \centering
    \includegraphics[width=0.5\textwidth]{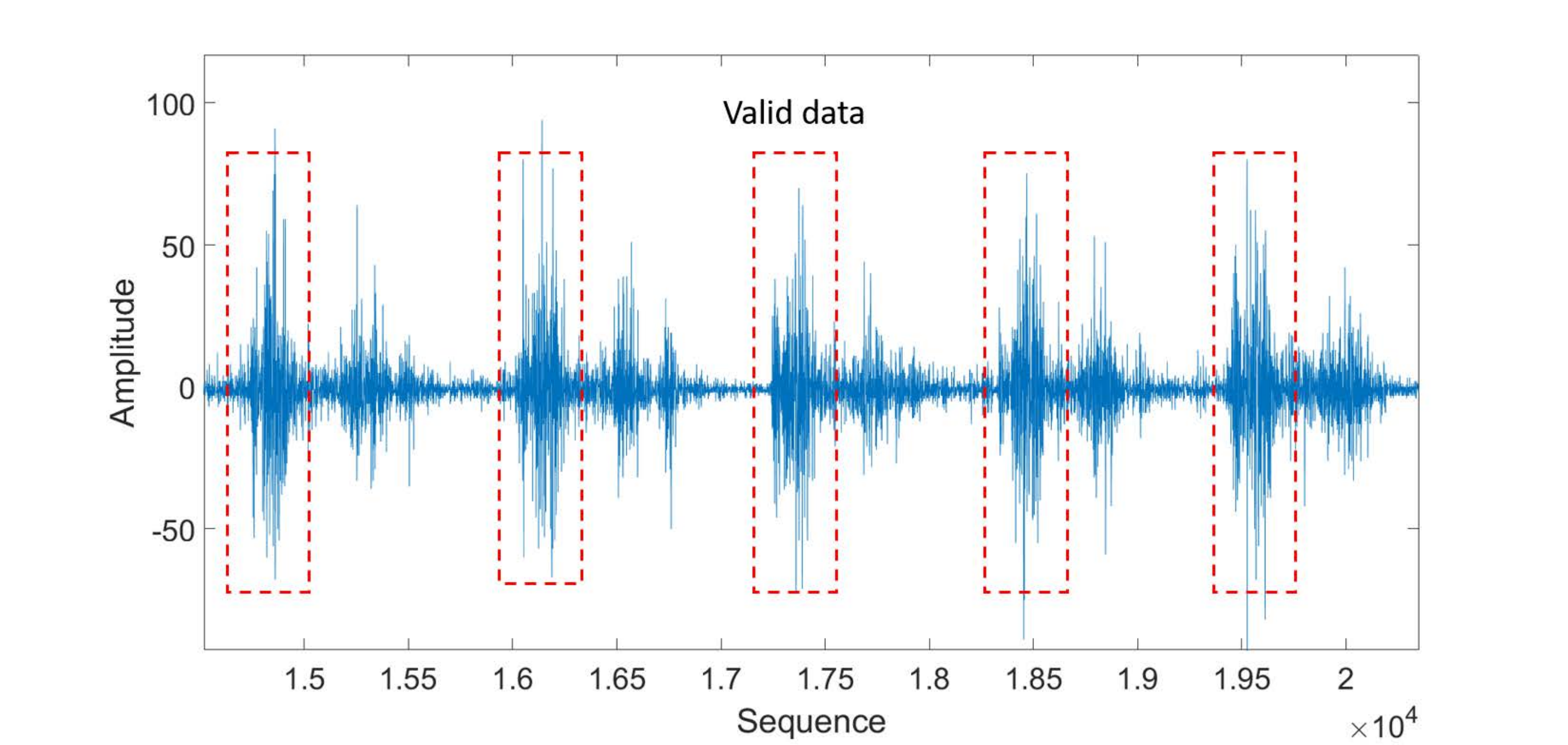}
    \caption{Examples of the raw data from the Myo armband.}
    \label{rawdata}   
\end{figure}  

\subsection{Data Segmentation}
In most cases, data collected by sensors should be denoised. Various filters are applied to make the signals smoother and more stable \cite{hines1999electronic}. In the popular EMG-based applications, such as gesture recognition, the signals should be roughly the same when different people do the same action \cite{wu2019design}. While in the area of user authentication, the tiny differences are crucial. EMG is the external quantified expression of the bioelectrical signals, which represents the structure of muscles and the amount of muscle contraction. Figure \ref{rawdata} presents the raw EMG signals from Myo, showing five similar partial waves which present five times of picking up and putting down the smartphone. The parts contained in the five red dotted line boxes are valid data and we need to extract them as the EMG matrices which will be described later. Here, we keep the raw signals without any filtering to maintain the uniqueness and use the convolutional network to extract features.   

We observe that the time from picking up the smartphone to watching the screen is generally no longer than two seconds. Hence, we set the valid action time to two seconds. As the sample rate is 200 Hz and the number of channels is eight, we take an $8\times400$ signal matrix as one piece of basic training data. It is comparable to a picture with its resolution being $8 \times 400$ rather than $32 \times 32$ or $1024 \times 1024$. We divide these matrices into different groups according to the corresponding people and make pairs according to the following rules: the label of a pair that comes from different people is set to 0 while the same person is set to 1. The pair construction algorithm is shown as Algorithm \ref{makepairs}. The data after pairing is fed into the neural network for training. 

\begin{algorithm}[h]
    \caption{Pair Construction Algorithm}
    \label{makepairs}
    \begin{algorithmic}[1]
        \State Input $Data$
        \State $len = length(Data)$
        \For{each $person\in data$} 
            \For{$i \in length(person)$}
            \State $pair1=Data[person][i]+Data[person][i+1]$
            \State $y(pair1)=1$
            \State $index = Random(0,\ len)$
            \State $dif\_person=(person+index)\ \%\ len$
            \State $pair2=Data[person][i]+Data[dif\_person][i]$
            \State $y(pair2)=0$
            \State $FinalData.append(pair1,pair2,y)$
        \EndFor
        \EndFor\\
        \Return $FinalData$
    
    \end{algorithmic}
\end{algorithm}

\subsection{Armband rotation-independence method}
\begin{figure}[H]
  \setlength{\abovecaptionskip}{-0.cm}
  \setlength{\belowcaptionskip}{-0.cm}
  \centering
  \subfigure[Roll one channel]{
    \label{roll1} 
    \includegraphics[width=0.4\textwidth]{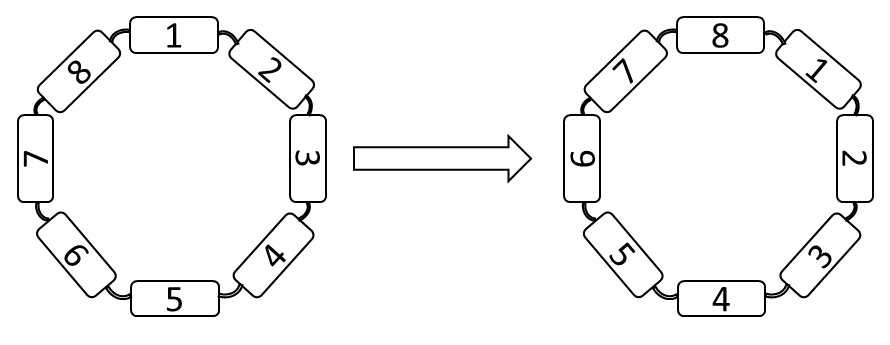}
  }

  \subfigure[Roll less than one channel]{
    \label{roll2}
    \includegraphics[width=0.4\textwidth]{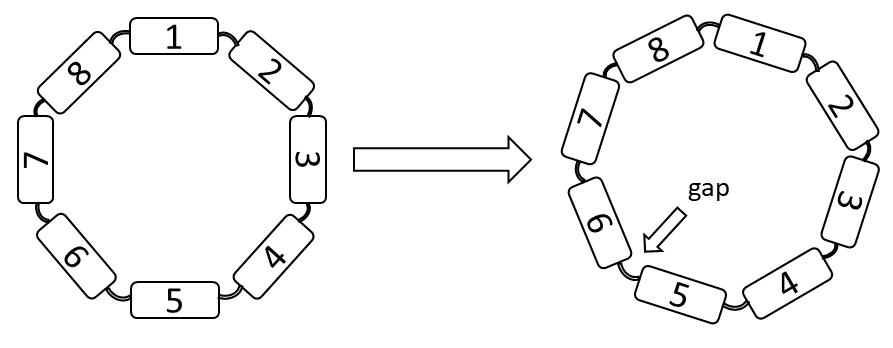}
  }
  \caption{Myo rotation sketch map.}
  \label{myotop}   
\end{figure}

One significant challenge that we need to address is the rotation problem of Myo armbands. There are eight EMG sensors in the armband and each sensor corresponds to a specific skin area. We can not rotate the armband freely because the EMG signal is unique among different skin areas. However, fixing the position means we have to mark the position or calibrate every time we wear it, which is inconvenient. To address this challenge, we propose a novel method based on the structure of Myo armband to make it rotation-independent. The Myo armband consists of eight rectangular sensors and they have unified sizes. Due to the fixed relative positions among these eight sensors, i.e., if the first sensor rotates to the position of the second, all the rest seven sensors will move in order and the last sensor will replace the position of the first sensor, we leverage the data augmentation technique from image classification tasks to expand our dataset. In our dataset, the eight channels correspond to eight sensors. Every time we roll the channels, a new dataset is created. Figure \ref{roll1} shows the result of rolling one channel. We get an eight times dataset until we roll a complete circle. 


From the Figure \ref{roll2}, we can see there is a gap between two sensors. When the user wears the armband, we cannot guarantee the position is just one of the eight positions that we expanded and the sensor may cover the gap area when the rotation is less than one channel. However, the distance of this gap is much less than the width of a sensor, the impact of the gap on EmgAuth is in turn limited. Rotation leads to many possibilities but we can only choose some representative positions to train the model. There is a trade-off between the accuracy and the computation complexity. Moreover, the probability that the sensor exactly falls on the gap is very small, and the system accuracy will increase as the area touched by the gap decreases.

In the task of image classification, flipping, cropping and scaling are the common data augmentation techniques \cite{perez2017effectiveness}. After these operations, the label of an image does not change. Our labeling process is similar, after channel exchange, the expanded dataset still belongs to one person. In this way, the deep neural network can learn enough features and make reliable decisions, no matter how the user wears the Myo armband and whether the user rotates it or not.

\begin{figure*}
    \setlength{\abovecaptionskip}{-0.cm}
    \setlength{\belowcaptionskip}{-0.cm}
    \centering
    \includegraphics[width=0.85\textwidth]{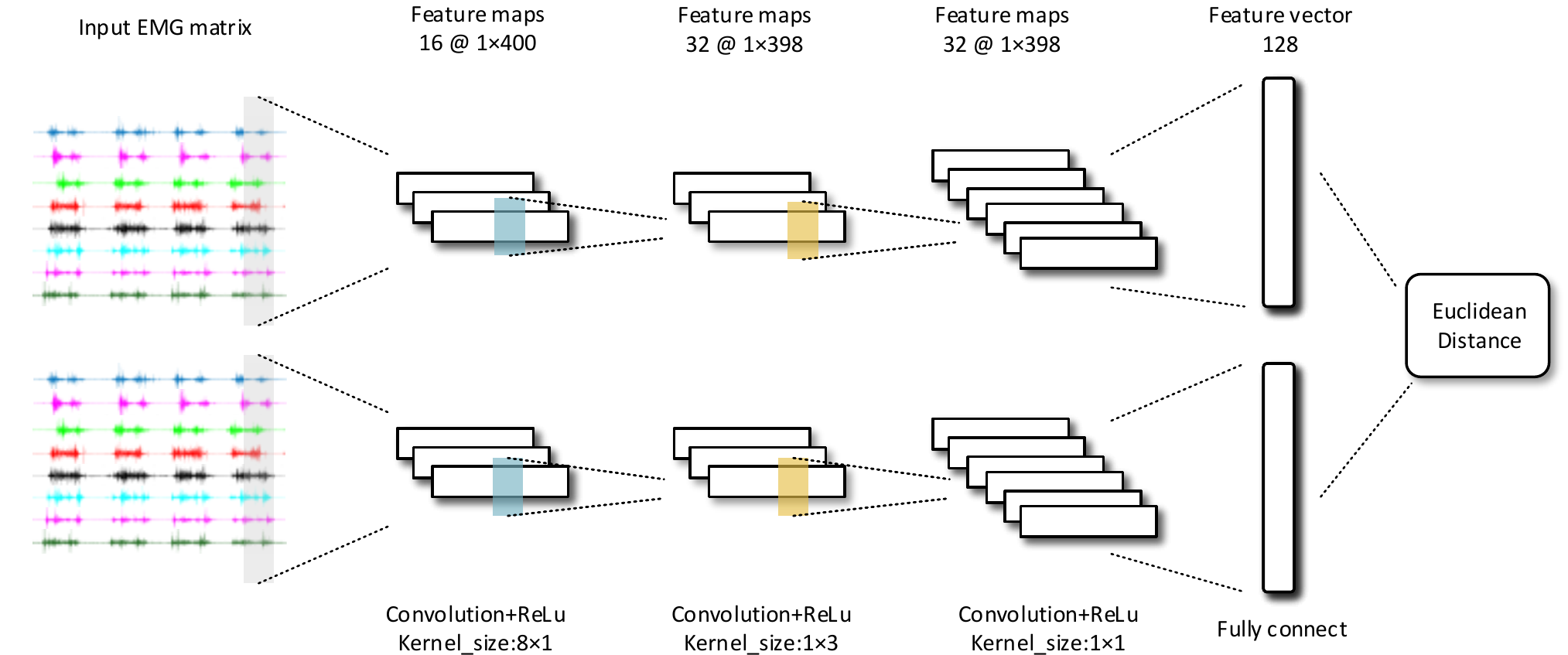}
    \caption{Structure of our convolutional Siamese network.}
    \label{network}   
\end{figure*}  

\subsection{Siamese network}
A standard CNN typically requires a large amount of data to train a robust model. While we use our own dataset, the amount is limited and each class has only 70 data points, which is far less than the empirical amount of data for CNN training. Besides, the users of authentication are dynamic as there are always users joining and leaving. However, retraining the model when user group changes introduces very high costs for data collection and periodical re-training. Considering these requirements, we select the Siamese network as our deep learning model. 

Siamese networks does not require too many instances of a class and only a few are enough to build a satisfactory model. Instead of calculating many probabilities and directly classifying an input data to one of the classes, the Siamese network takes an extra data of the person as input and will produce a similarity score denoting the chances that the inputs belong to the same person. To be more specific, it has two inputs and one output whose value corresponds to the similarity between the two inputs. This network consists of two identical sub-networks with the same layers and weights. In addition, we add a layer to calculate the distance of the outputs of these two sub-networks, which will be used to compare with the threshold to decide the authentication result. 

For the type of sub-networks, we choose CNN as it can achieve extraordinary performance in local feature extraction. In our application, EMG signals are collected by eight sensors and we need to obtain the features of a single one and the combination. Therefore, we design a convolutional Siamese network and the architecture is shown in Figure \ref{network}. Our network architecture consists of three convolutional layers with different numbers of filters and one fully connected layer with 128 units. 

Considering the Myo armband returns eight channels' signals at the same time, we need to combine the data of these. Convolution operations can achieve this by sliding the convolution kernel. In the first convolutional layer, we set the kernel size to $8 \times 1$ to learn features among eight channels. We set the stride to one so that the first convolutional layer can focus on finding features among different channels. Since our input is an $8 \times 400$ matrix, the output size of the first layer is $1 \times 400$, which achieves the combination of eight channels. Next, we set the kernel size to 1 $\times$ 3 to extract features during the process of picking up a smartphone and we get a size of $1 \times 398$ in this layer. To reduce the number of feature maps, we add a convolutional layer with 1 x 1 filters. These take all features from the previous layers into the next fully-connected layer. Besides, we add dropout layers after each convolutional layer to prevent overfitting. The dropout rate is increasing with the depth of the network from 0.1 to 0.2. As for activation, we use Rectified Linear Unit (Relu) for nonlinear transformation. Relu can reduce the likelihood of vanishing gradient, which results in faster learning.

In the last two layers of the network, we use a flatten layer and a fully-connected layer. The flatten layer is used to flatten the output of previous convolutional layers so the features can be fed into a fully-connected layer. The fully connected layer takes every combination of features from the outputs of previous layers into account. Here, we do not have a softmax layer as usual since we prefer a vector that represents the original EMG input rather than a classification possibility. We define the number of units to 128, so we could get a 128-length vector as the map of the input EMG signals.

After defining the sub-network of our Siamese network, we need an extra layer to combine the outputs of them. In this layer, we use the Euclidean distance to measure the differences between two output vectors from the last two fully connected layers. Loss function is used in supervised machine learning to minimize the differences between the predicted output of the model and the ground truth labels. In our task, we use the contrastive loss to train our model. This loss function encourages the neural network to learn an embedding to place samples with the same labels close to each other, while distancing the samples with different labels in the embedding space.

We train the network to make the distances of data from different participants be as far as possible, while from the same participant be as close as possible. In the long run, the network will learn to extract meaningful features and has the ability to distinguish different people. The input shape is (8, 400, 1), where 8 means the data has eight different channels, 400 is the valid signal length, and 1 means each cell of the signal matrix has only one value.

\begin{figure}[!htbp]
  \centering
  \begin{minipage}[t]{0.2\textwidth}
      \centering
      \includegraphics[width=0.85\textwidth]{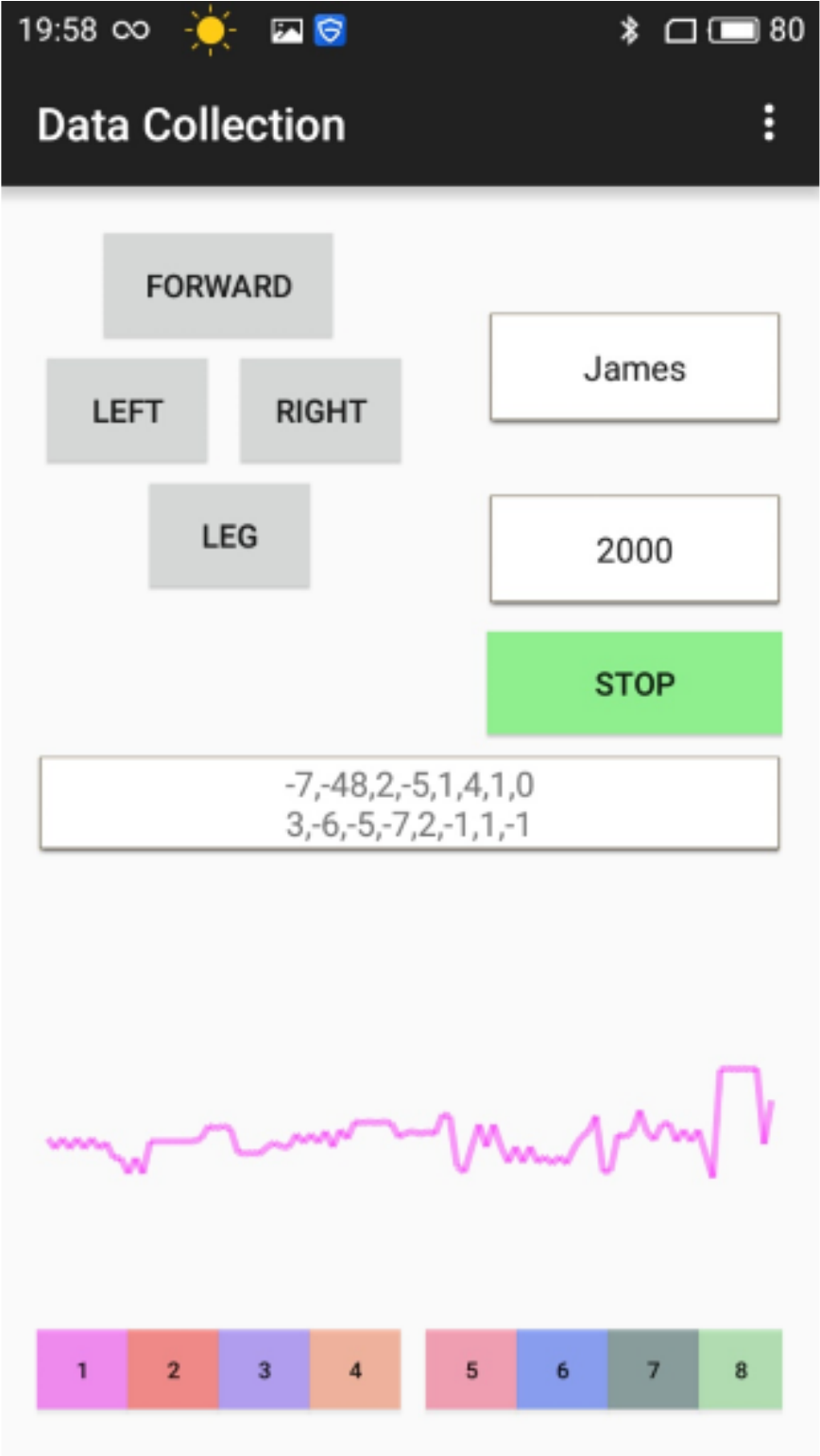}
      \caption{Data collection application.}
      \label{dataapp}   
      \end{minipage}
  \begin{minipage}[t]{0.2\textwidth}
      \centering
      \includegraphics[width=0.85\textwidth]{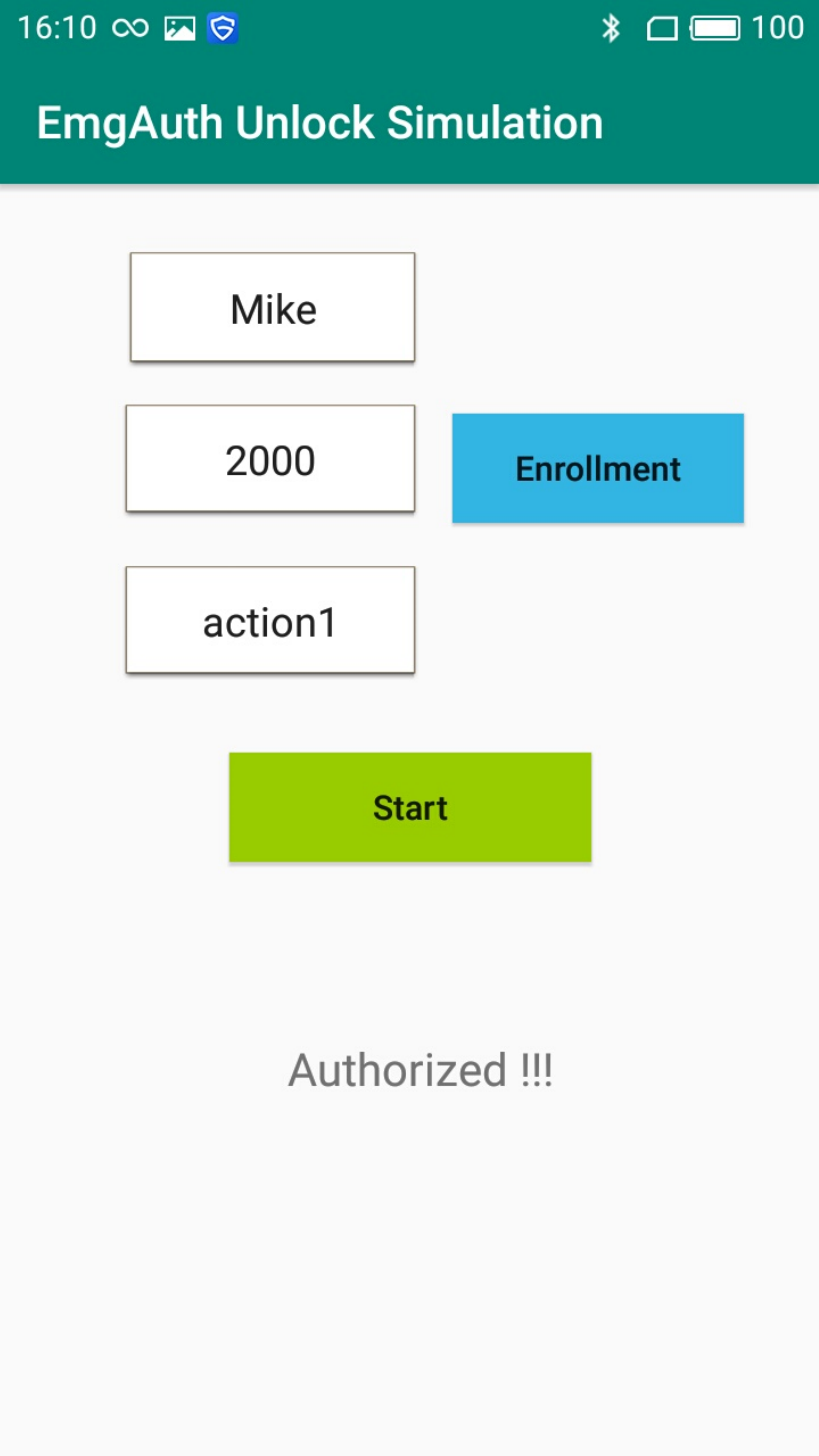}
      \caption{Simulation application.}
      \label{siamulation}   
      \end{minipage}
\end{figure}

\subsection{Unlock Prototype System}
To prepare the training data and make EmgAuth easy to use, we have developed a prototype system on Android, which can perform data collection and unlocking simulation.

\subsubsection{Data Collection Interface}
The data collection interface is shown in Figure \ref{dataapp}. There are four action buttons in the upper left corner, for example, left means the smartphone is placed to the left. The two input boxes in the right are used to input the the participant's name and the corresponding sampling time. The sampling time is the time from pressing the corresponding action button to the end of the action. During this time, the system will tag the EMG signal with the corresponding action label. The middle box shows the real-time EMG signals in numeric form and the lower area presents the real-time signal graph. The collected data will be saved in the format of a CSV file to facilitate data preparation for model training.
In the process of data collection, due to the possibility of some fluctuations in the sampling capacity of the equipment, the application can fix the length of the data to 400 automatically, adding 0 to the length is less than 400, and randomly deleting data, i.e., when over 400.

\subsubsection{Unlocking Simulation Interface}
Considering that smartphones have two primary states before it is used, i.e., horizontal placement (for sitting) and vertical placement (for standing), the system should automatically select the model according to the state of the smartphones. This can be achieved by using the built-in accelerometer. The accelerometer has three axes, namely, the x-, y-, and z-axis. When a smartphone is horizontal, the absolute value of z-axis is the largest; when the phone is vertical, z-axis has the minimum value. Therefore, the system can easily choose the proper model to load based on this rule.

We transfer the fine-tuned models to the smartphones. Figure \ref{siamulation} presents the user interface of the unlocking simulation application. As the first step, the user needs to input his or her name as the index of later EMG signals. In the second step, the user performs different actions to save their unique EMG signals in the database. We name these two steps as the enrollment stage. In the authentication stage, users open the simulation interface and take their smartphones as normal. The application saves the EMG signals from the Myo armband in real-time and implements segmentation in the time window of two seconds, so the processed signals can be fed into our model. The signals after segmentation are then paired with the previously stored different types of EMG matrices, respectively. These pairs are fed into the model, if one of the output is less than the threshold, the system considers the authentication successful; otherwise, the newcomer user is rejected.

\section{Experiment and Evaluation}
This section presents the experiments and implementation details of the EmgAuth system. We collect data from 53 participants and use this dataset to train a neural network. We then present the training process and show the influence of related parameters. The performance of EmgAuth is tested in several experiments as well as the impact of different factors. We also verify the rotation-independence feature through additional experiments.

\subsection{Dataset}

In order to collect data to train a deep learning model, we invite 53 participants to help us build an EMG signal dataset as well as validation sets for testing rotation-independence and non-sitting scene. Participants include 32 males and 21 females, with the mean age of 25.3 and a range from, 18 to 47, which is considered to be typical for user groups of smartphones. 

\begin{figure}[h]
  \setlength{\abovecaptionskip}{-0.cm}
  \setlength{\belowcaptionskip}{-0.cm}
  \centering
  \includegraphics[width=0.2\textwidth]{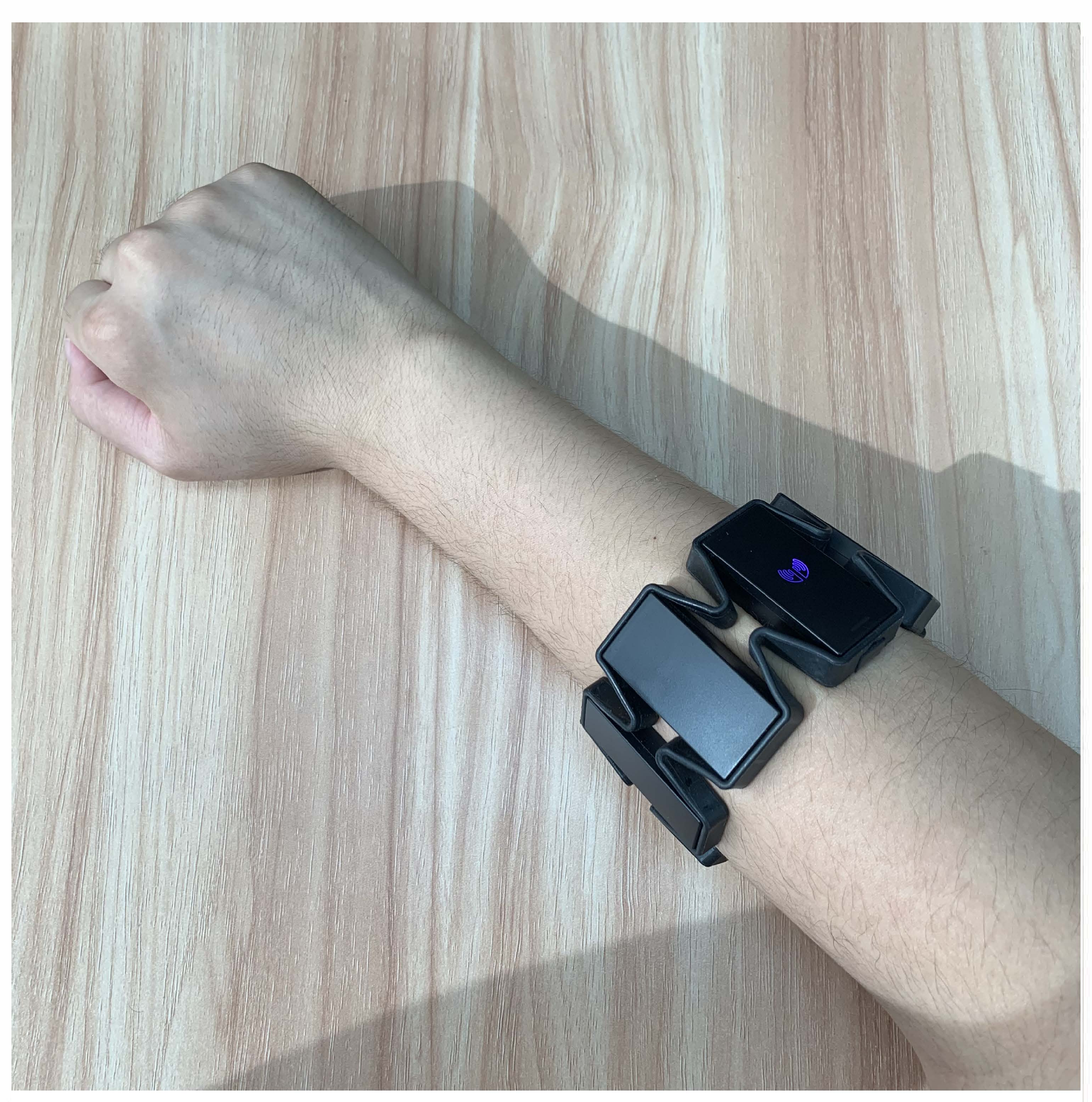}\\
  \caption{The wearing location of the Myo armband.}
  \label{armband}   
\end{figure}

\begin{figure}
  \setlength{\abovecaptionskip}{-0.cm}
  \setlength{\belowcaptionskip}{-0.cm}
  \centering
  \includegraphics[width=0.28\textwidth]{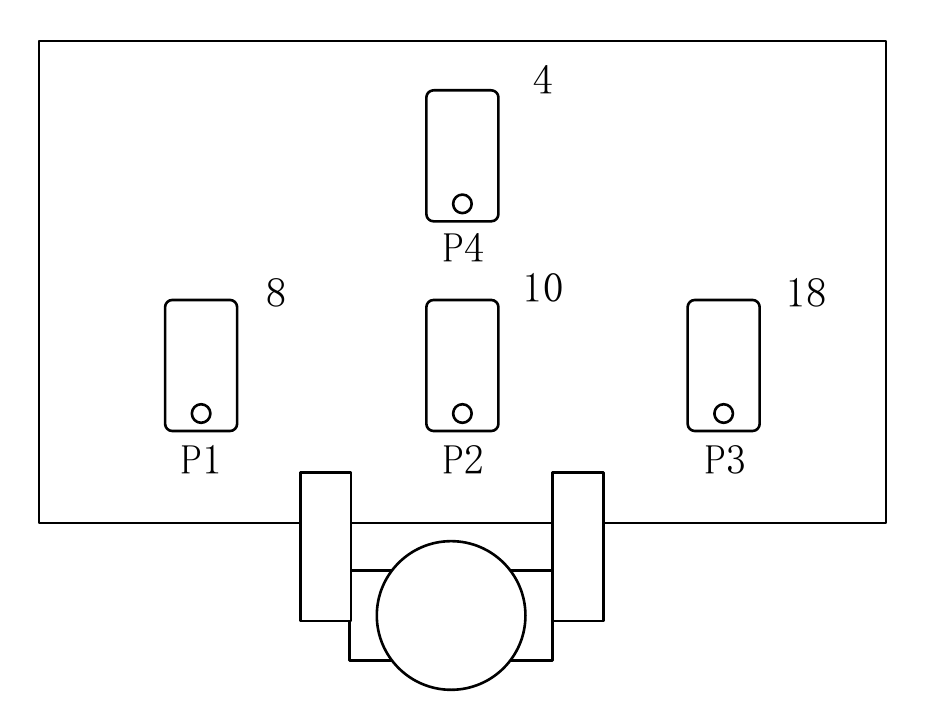}\\
  \caption{Smartphone positions and corresponding frequencies.}
  \label{desk}   
\end{figure}

We split the data from the 53 participants into a training dataset of 40 participants for training, and 13 for testing and validation. They are required to wear the Myo armband on the forearm. The position of Myo armband is shown in Figure \ref{armband}. This process is conducted in an academic environment and all participants are free from being disturbed. An Android smartphone runs our data collection application instead and the EMG signals are recorded into CSV files. We design a questionnaire to record smartphone positions when people are studying or working, the results are presented in Figure \ref{desk}. As we can see, people usually place their smartphones on the left, forward, and right when they are sitting, this helps us to design the data collect motions.

\begin{figure}
  \centering
  \subfigure[Step 1 (hand on the leg)]{
    \includegraphics[width=0.2\textwidth]{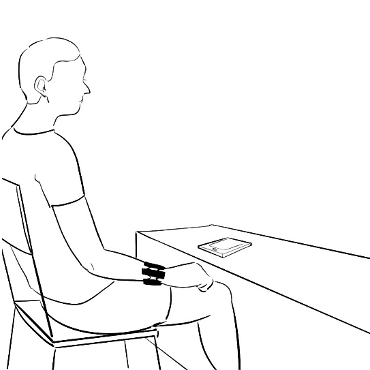}
    \label{step11}
  }
  \subfigure[Step 1 (hand on the table)]{
    \includegraphics[width=0.2\textwidth]{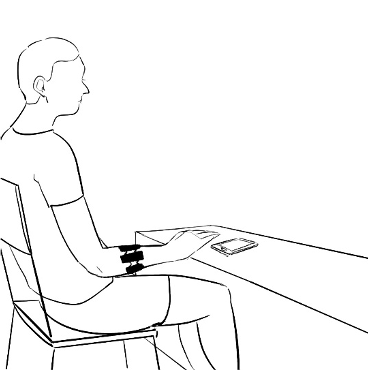}
    \label{step12}
  }
  \subfigure[Step 2 (pick up the phone)]{
    \includegraphics[width=0.2\textwidth]{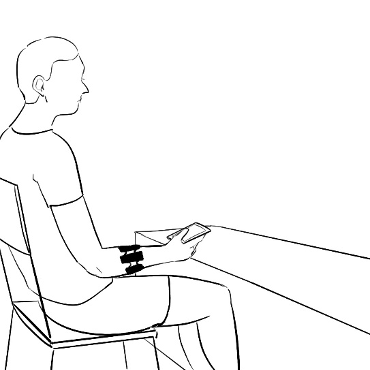}
  }
  \subfigure[Step 3 (watch the screen)]{
    \includegraphics[width=0.2\textwidth]{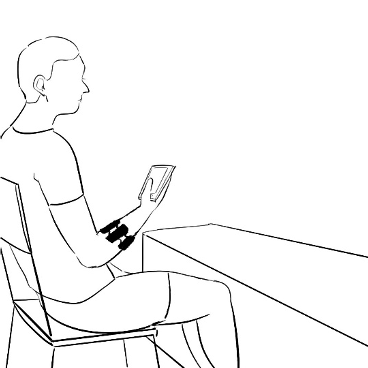}
  }
  \caption{Data collection steps.}
  \label{datacollection}   
\end{figure}

We require the participants to sit in front of a desk and place their smartphones on the desk. As shown in Figure \ref{datacollection}, the motion they need to do is just picking up their smartphones and watching the screens as usual. Here, we design four scenes and the only difference among them is the location of the smartphone and hand. There are two initial positions for hand, which are shown in the Figure \ref{step11} and \ref{step12}. There is just one smartphone position corresponding to the initial action while three positions for the latter initial action. The four smartphone positions are shown in Figure \ref{desk}.

In position P1, P2, and P3, participants are required to repeat the motion twenty times while for position P4, ten times. As a result of our survey, we decrease the number of times for the last scene because most people do not place their smartphones in such a position. The valid time starts from the user beginning to move his or her hand to look at the screen and this time is normally less than two seconds, we therefore set the valid time to be the same amount of time. To be specific, the sample rate is 200 Hz and the Myo armband has eight channels, so one piece of valid data is an $8 \times 400$ matrix. With 40 participants, we collect a dataset with 2800 pieces of valid data.

We ask the other 13 participants to help us build two datasets for rotation-independence verification and non-sitting scene verification. To collect data for the first dataset, participants are required to do the following steps. Firstly, they wear the Myo armband on the right arm, record the wearing position at this time, and perform the pick-up action ten times. Secondly, they rotate the armband clockwise by one sensor position, then do the same action ten times. Thirdly, they rotate four sensor positions clockwise, do the same action ten times. Finally, they rotate the position to the gap, and record the wearing position at the same time, and perform the pick-up action ten times. According to the previous data processing method, we collect a small verification dataset of 520 pieces of data. To collect data for the second dataset, participants conduct the following, wearing the Myo armband on the right arm, hold the smartphone in right hand, then naturally raise their hand to look at the phone and then place it down, they repeat this action ten times. A validation set with 130 valid data points can be obtained and then used to train a tiny Siamese network to adopt the non-sitting scene.

\subsection{Model Training and Evaluation}
We implement our model using Keras, a Python-based deep learning platform. We train our model on a server machine equipped with an NVIDIA Tesla V100 GPU, 128 GB memory, and an Intel Xeon E5 2560 processor. We use an Adam optimizer with a learning rate of 0.002 and a batch size of 32. For loss function, we choose contrastive loss rather than cross entropy loss. Contrastive loss runs over pairs of samples. During training, an EMG signal matrix pair is fed into the model with their ground truth relationship $Y$. If the two matrices are similar, $Y$ equals 0; otherwise, $Y$ equals 1. The loss function is defined as ~\eqref{loss}, where \textit{d} is the Euclidean distance between the two EMG feature vectors. The margin term is used to keep the loss within a valid range. For example, if two EMG signal matrices in a pair are dissimilar, then their distance should be at least the value of margin, otherwise the loss will be 0.

\begin{figure}
  \setlength{\abovecaptionskip}{-0.cm}
  \setlength{\belowcaptionskip}{-0.cm}
  \centering
  \includegraphics[width=0.48\textwidth]{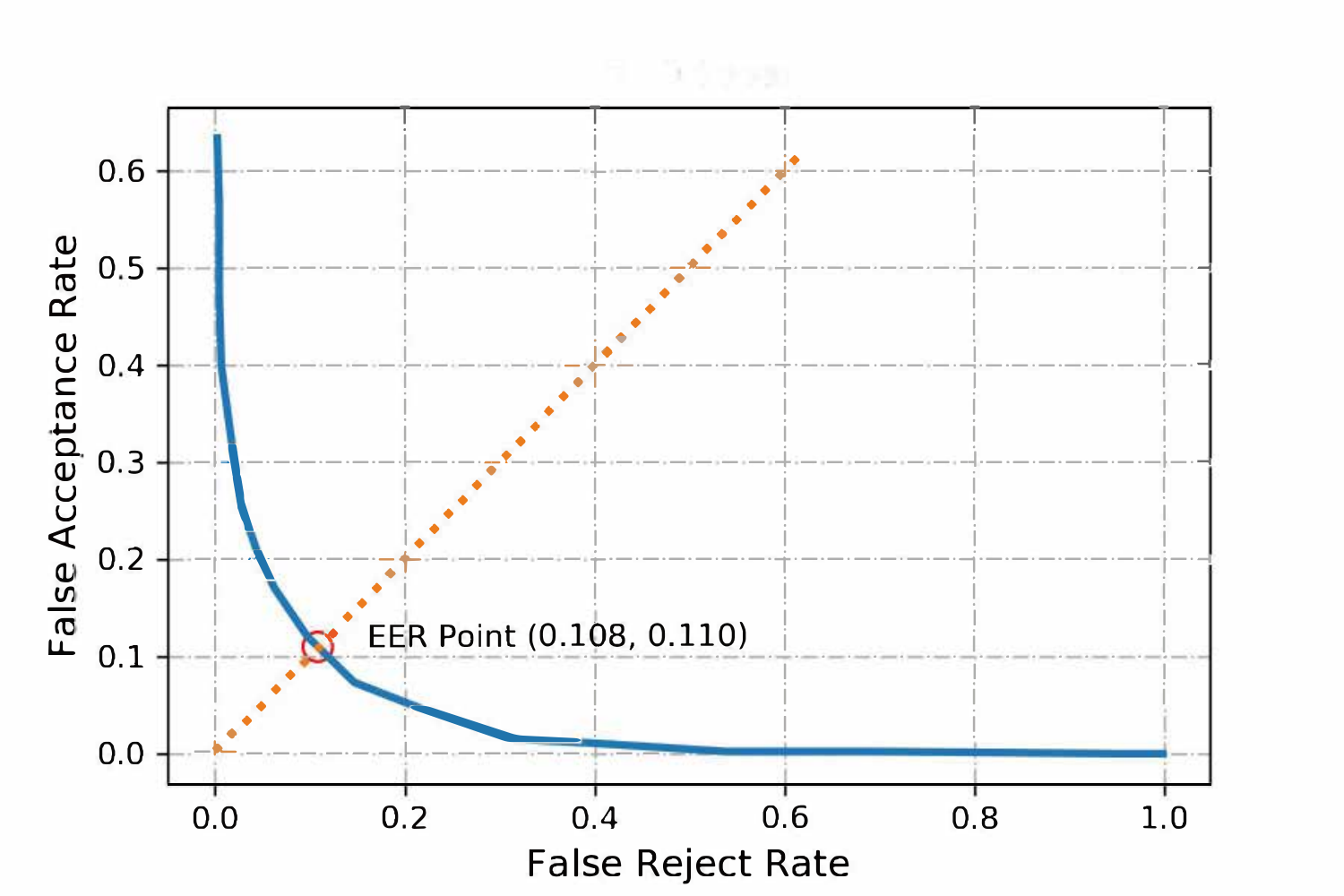}
  \caption{DET curve with different thresholds.}
  \label{det}   
\end{figure}  

\begin{equation}\label{loss}
    Loss = (1-Y)d^{2} + Y\left \{ max(0,margin - d) \right \}^{2}.
\end{equation}


We evaluate the performance of EmgAuth using metrics that are commonly used in evaluating authentication systems. These metrics include accuracy, true acceptance rate (TAR), false acceptance rate (FAR) and false rejection rate (FRR). Equal error rate (EER) is also leveraged to find an appropriate classification threshold. With threshold increases, FAR drops while FRR increases. EER is the point that FAR equals with FRR. The classifier has the best performance when the threshold corresponds to EER. Hence, we plot the Detection Error Tradeoff (DET) curve and the result is presented in Figure \ref{det}. The corresponding threshold of this EER point is 0.55, so we use it as the final threshold of our system.



We then study the effect of the different hyperparameters on the system performance, including the number of CNN layers, filter shape, learning rate, dropout rate, batch size, and the number of epochs. Table \ref{parameter} shows the initial hyperparameters used in the evaluation section. We use the hold-out validation to check the performance and the ratio of train set and test set is 4:1. Also, we use accuracy as the only metrics of this evaluation.


\begin{table}[]
  \centering
  \caption{Training parameters}
  \label{parameter}
  \begin{tabular}{@{}lllll@{}}
  \toprule
  Paremeter            & Range         & Initial Value &  &  \\ \midrule
  Number of CNN layers & 2 - 6         & 2             &  &  \\
  Filter number        & 16 - 64       & 16            &  &  \\
  Dropout rate         & 0.1 - 0.3     & 0.1           &  &  \\
  Batch size           & 16 - 256      & 16            &  &  \\
  Number of epochs     & 10 - 50       & 10            &  &  \\
  Learning rate        & 0.001 - 0.005 & 0.001         &  &  \\ \bottomrule
  \end{tabular}
\end{table}

\begin{figure}
    \centering
    \subfigure[CNN layers]{
    \label{layers} 
    \includegraphics[width=0.23\textwidth, height=1in]{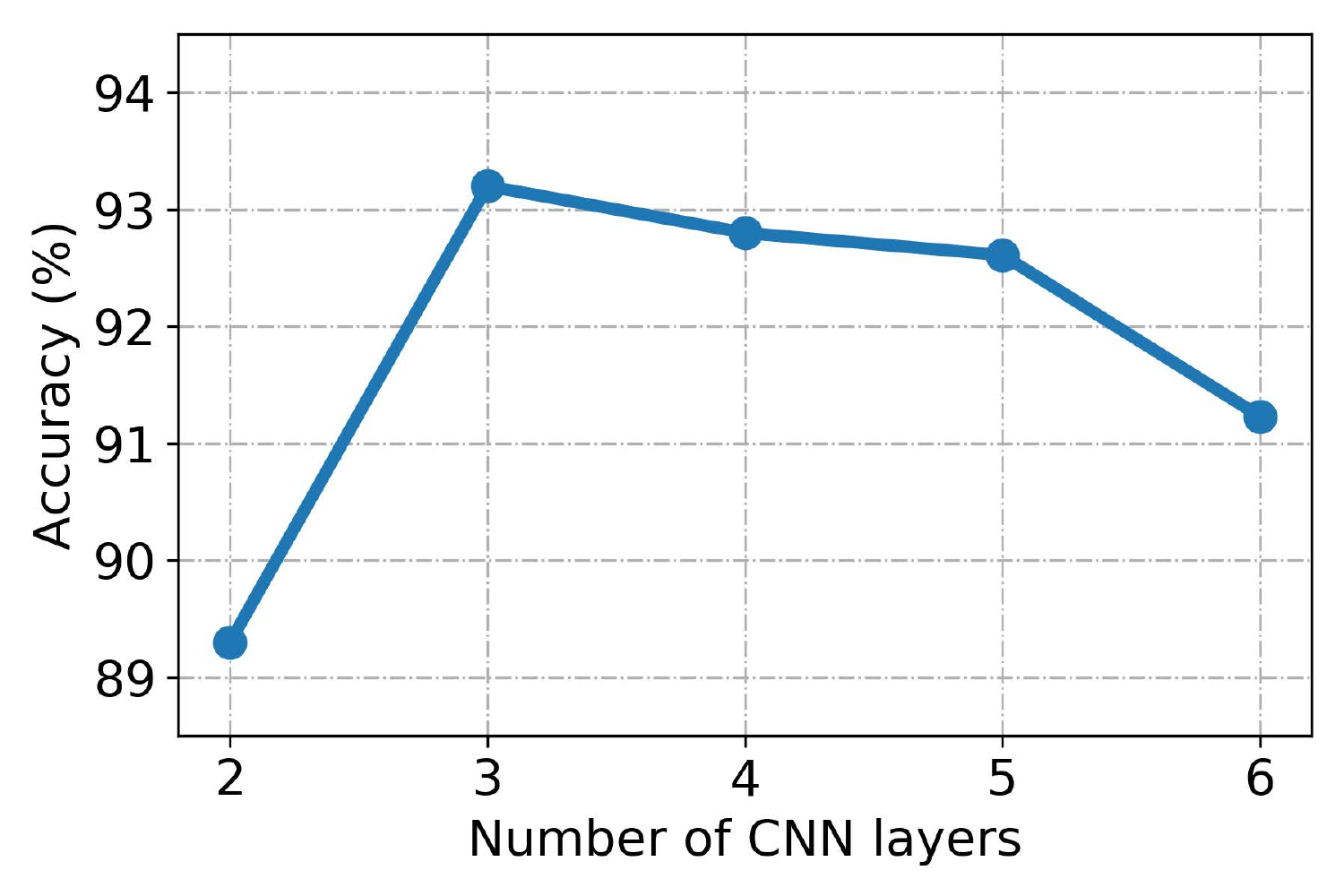}}
    \subfigure[Number of convolution filter]{
    \label{filter} 
    \includegraphics[width=0.23\textwidth, height=1in]{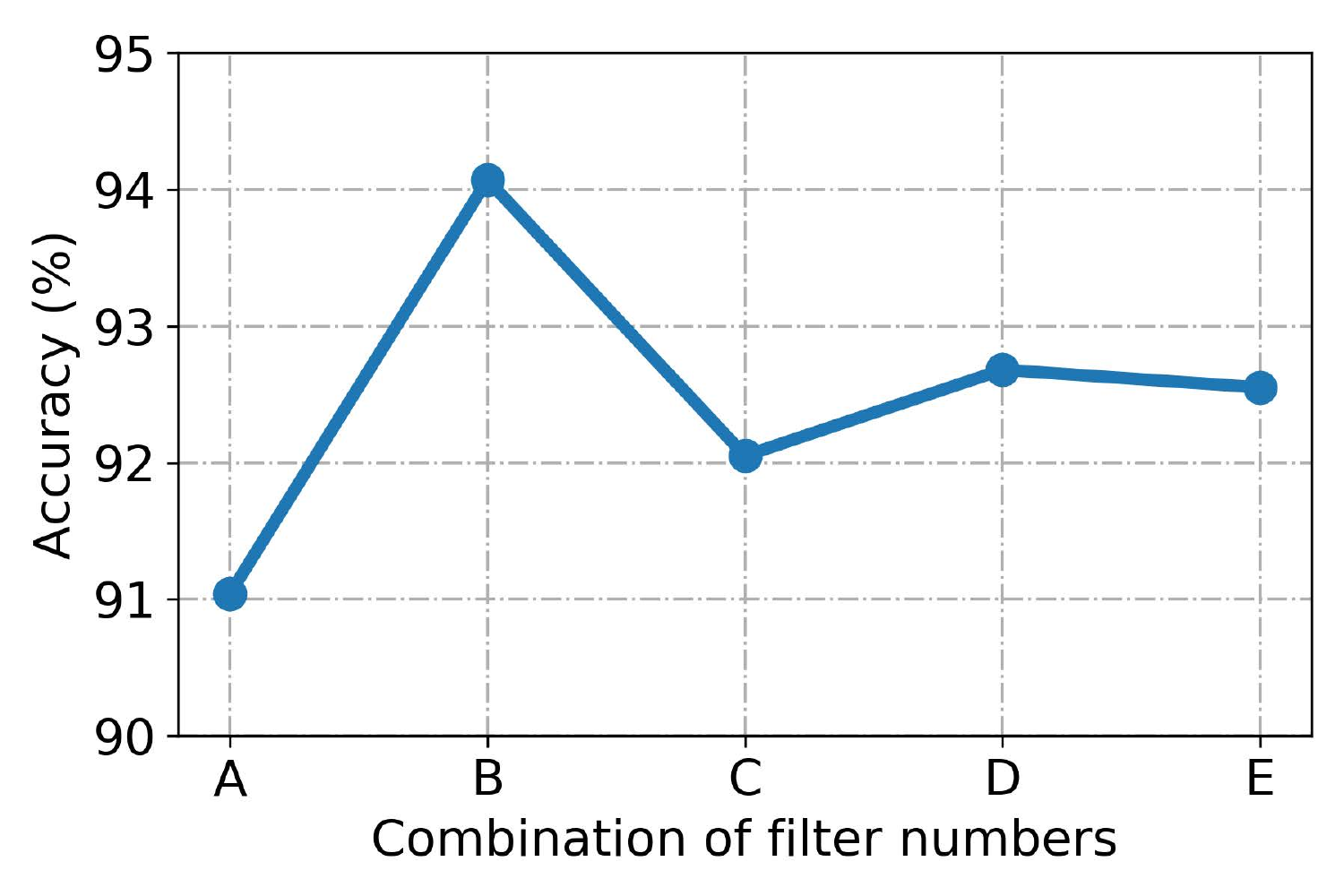}}
    \subfigure[Learning rate]{
    \label{learningrate} 
    \includegraphics[width=0.23\textwidth, height=1in]{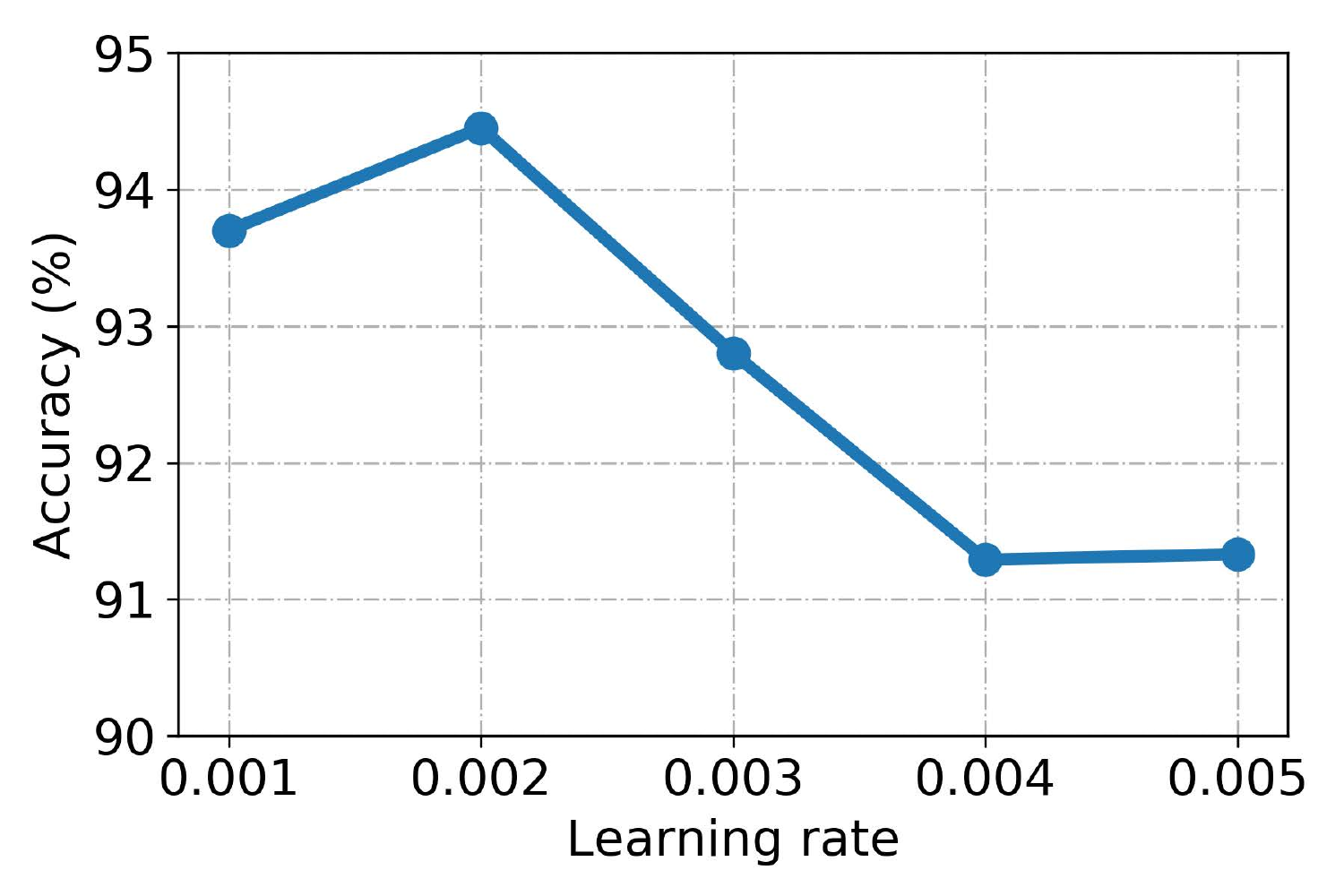}}
    \subfigure[Dropout rate]{
    \label{dropout} 
    \includegraphics[width=0.23\textwidth, height=1in]{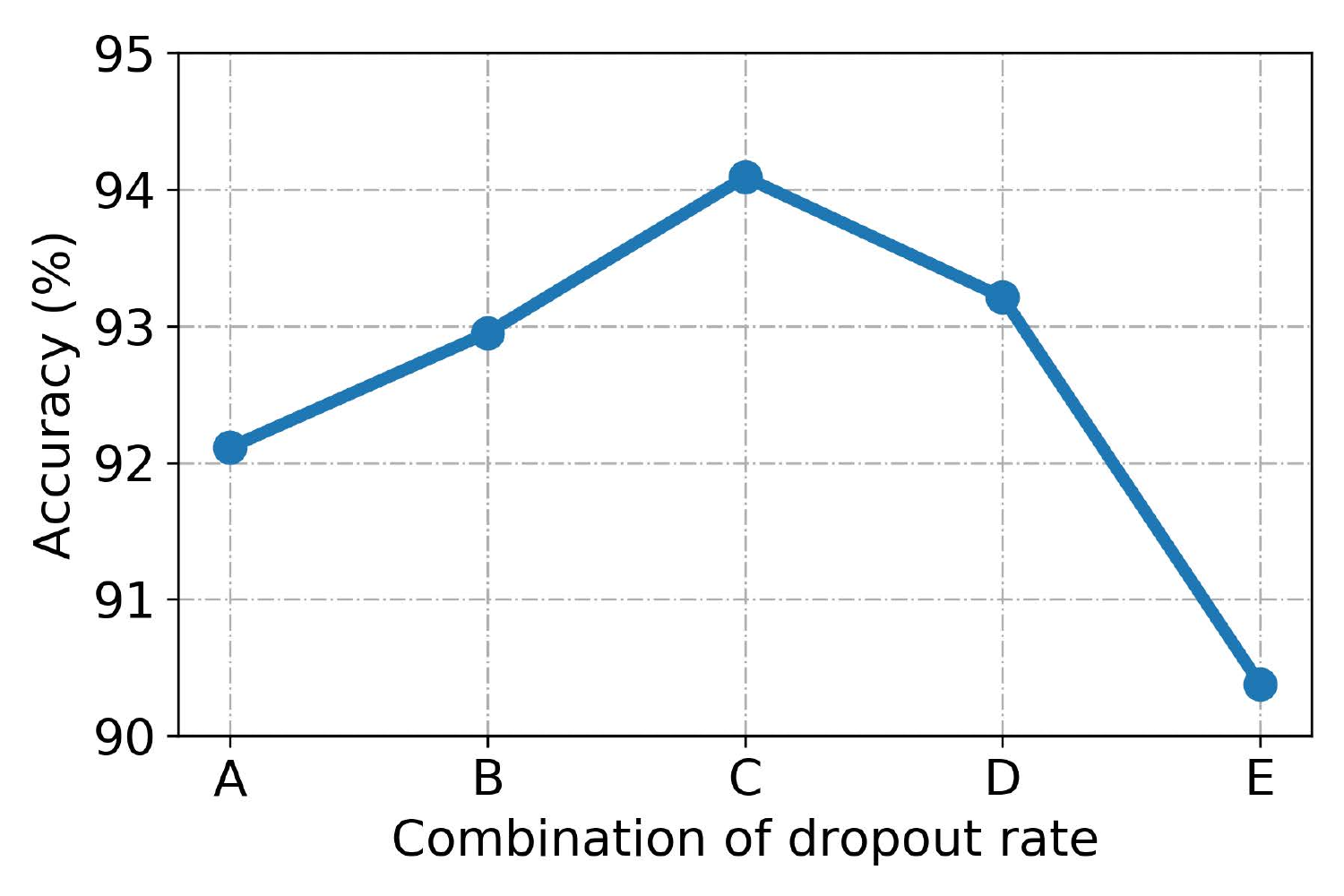}}
    \subfigure[Batch size]{
    \label{batchsize} 
    \includegraphics[width=0.23\textwidth, height=1in]{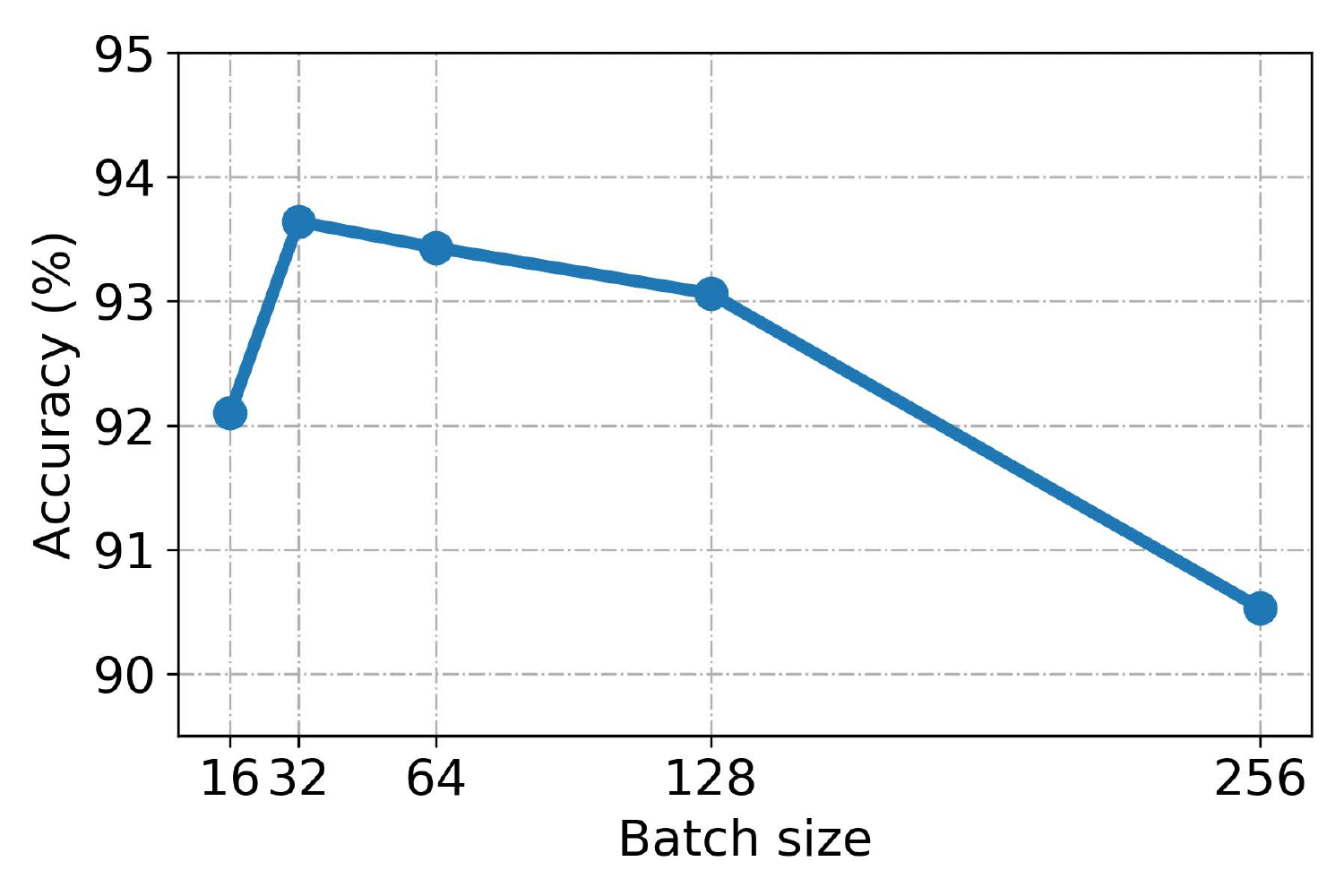}}
    \subfigure[Number of epochs]{
    \label{epoch} 
    \includegraphics[width=0.23\textwidth, height=1in]{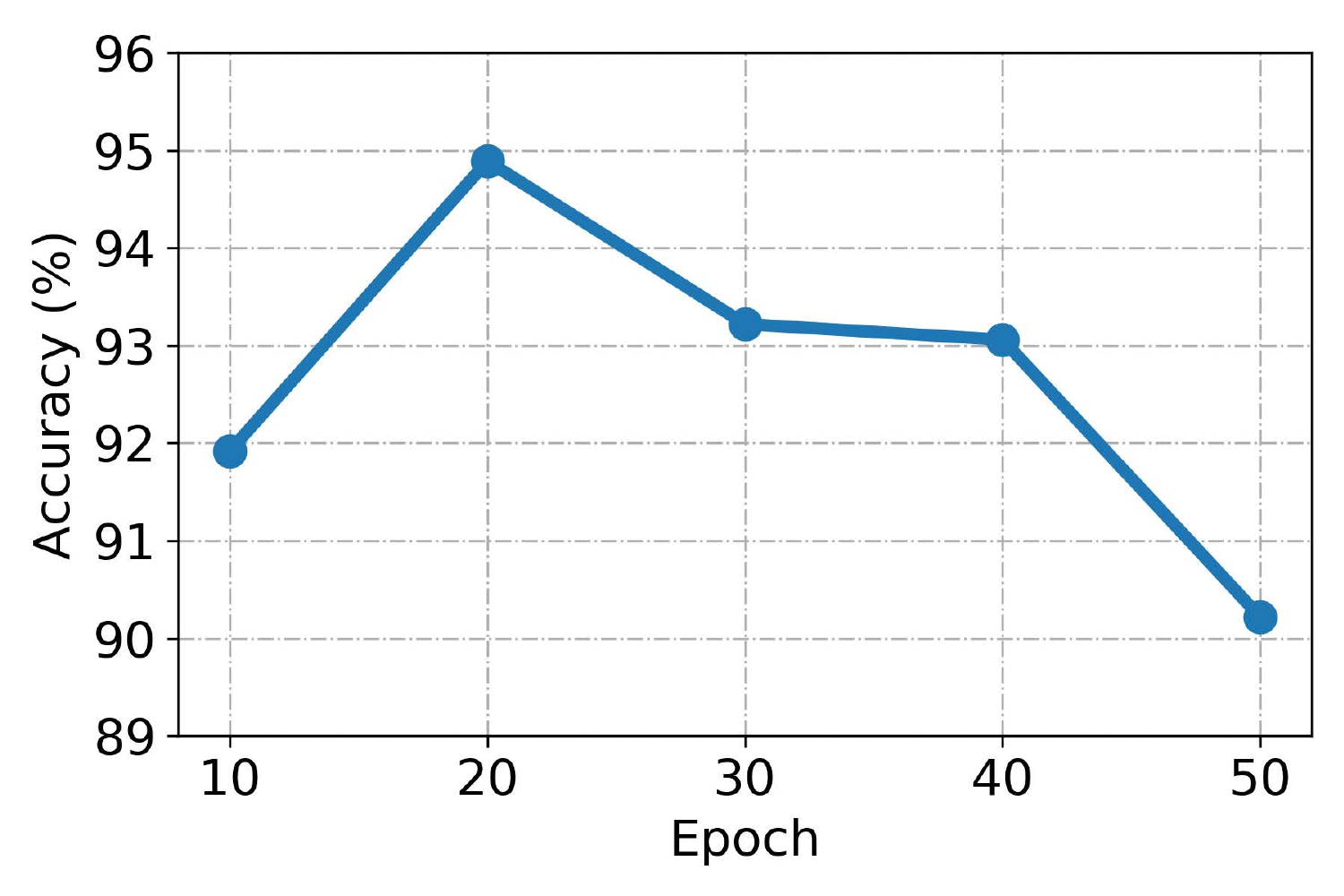}}
    \caption{Effect of different hyperparameters on authentication accuracy.}
    \label{allpara} 
\end{figure}

\subsubsection{Effect of the number of CNN layers}
Figure \ref{layers} presents the effect of changing the number of layers. The system reaches the highest accuracy when the number of layers is 3. We find as the number of layers increases, the accuracy drops, which is because the EMG signals are not as complex as the images. With too many layers, this may lead to overfitting and only two layers are not sufficient to learn enough features. Three layers are therefore the best choice for our task.

\subsubsection{Effect of different filter shapes}
Figure \ref{filter} shows the effect of the number of convolution filters in each layer. A suitable number of convolution kernels can fully extract the features of the signals. We try five different number of filter combinations because there are three convolutional layers. The A, B, C, D and E lettering corresponds to [16,16,16], [16,32,32], [16,32,64], [32,32,32] and [32,32,64], respectively. Combination B performs best, so we choose [16,32,32] as the filter number in the three convolutional layers. 

\subsubsection{Effect of different learning rates}
The learning rate is a hyperparameter that controls how much to change the model in response to the estimated error each time the model weights are updated. Too big or too small learning rates can both have negative effects on the learning result, for example, resulting in the model to not converge or train too slowly. Therefore, we use a series of values to choose the best hyperparameter. Figure \ref{learningrate} shows that 0.002 is the most appropriate learning rate. 

\subsubsection{Effect of different dropout rates}
Dropout is the simplest way to prevent a neural network from overfitting. Considering that each convolutional layer has a dropout layer, we try different dropout rate combinations and name them as in Figure \ref{filter}. Here, the A, B, C, D, and E correspond to [0.1,0.1,0.1], [0.1,0.2,0.2], [0.1,0.2,0.3], [0.2,0.2,0.2] and [0.2,0.2,0.3]. Figure \ref{dropout} shows that the dropout rate combination [0.1,0.2,0.2] is the most desirable.

\subsubsection{Effect of different batch sizes}
Figure \ref{batchsize} shows how the batch size affects our system. The batch size is a hyperparameter that defines the number of samples to work through before updating the internal model parameters. Considering that the size of our dataset is not very large, a big batch size is not a good choice. From the figure, we see that the accuracy reaches the highest when the batch size is 32.

\subsubsection{Effect of the number of epochs}
Figure \ref{epoch} presents the effect of training epochs. The number of epochs is the number of complete passes through the training dataset. If the number of epochs is too high, the model is easily overfitted as we have a small dataset. From the experiment, the model has the best accuracy when we set the number of epochs to 20. 

\subsubsection{Effect of different distance functions}
\begin{figure}
  \setlength{\abovecaptionskip}{-0.cm}
  \setlength{\belowcaptionskip}{-0.cm}
  \centering
  \includegraphics[width=0.48\textwidth]{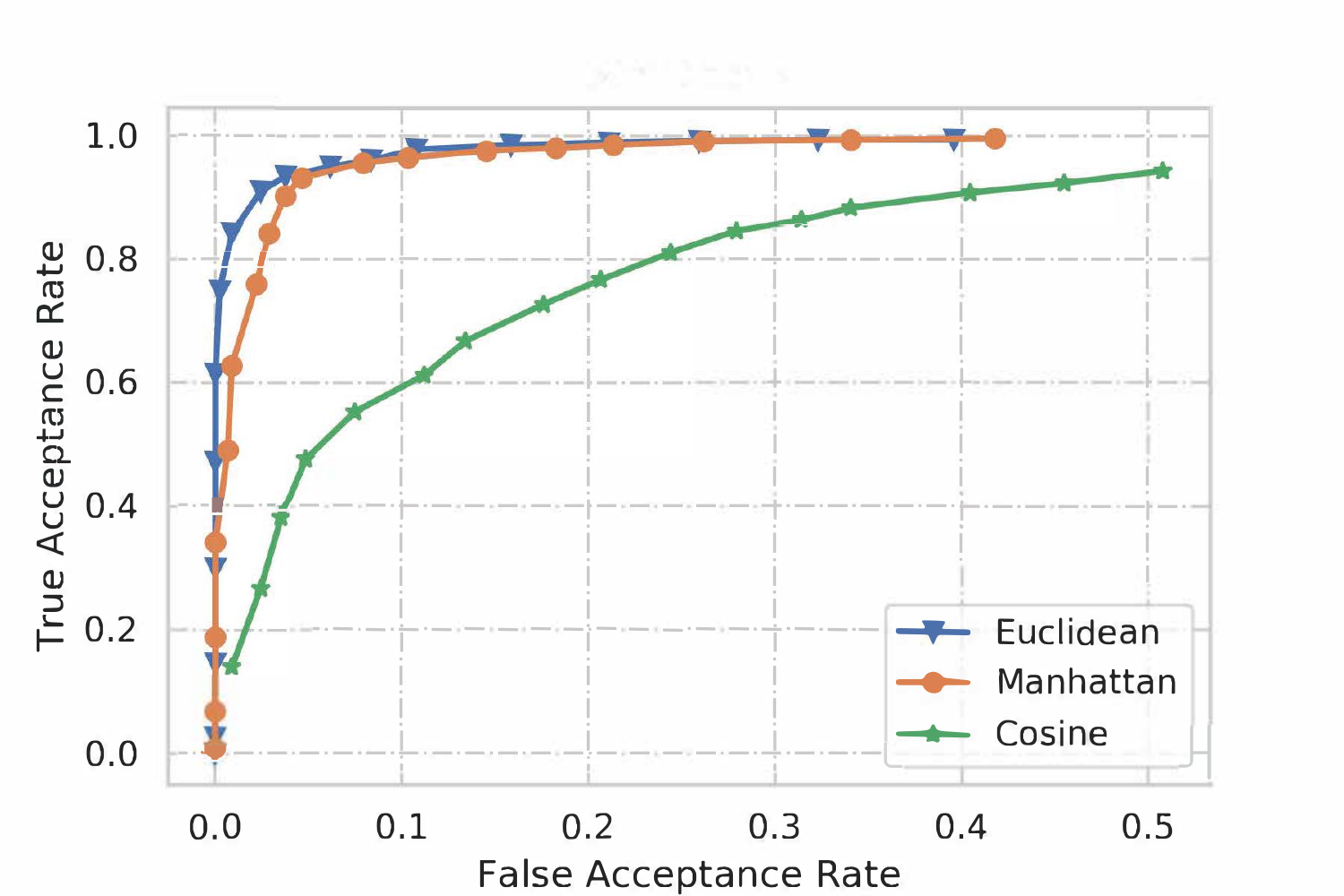}\\
  \caption{ROC curves of different distance functions.}
  \label{roc}   
\end{figure}

As mentioned above, we add a distance layer to combine the outputs of two identical sub-networks for measuring the similarity of them. Here, we evaluate three different functions including Manhattan distance\cite{greche2017comparison}, Euclidean distance, and Cosine distance\cite{george2015cosine} to find the best option.
To take a closer look at the system performance under different distance functions, we leverage the Receiver Operating Characteristic (ROC) curve in the study. ROC curve is an effective method to graphically reflect and compare the performance of different classifiers. Each point on an ROC curve corresponds to a certain detection threshold. Figure \ref{roc} presents the ROC curves of different distance functions, the Area Under the ROC Curve (AUC) of Euclidean distance is the biggest while Cosine distance performs worst. The AUC of Manhattan distance is between the other two distance functions. Therefore, we use Euclidean distance to measure the similarity of two feature vectors of the EMG signals.

After these experiments, we list the best parameters in Table \ref{final}. We use our fine-tuned network to evaluate the performance of the EmgAuth system. We divide our dataset into five subsets, which means each has 8 participants' EMG signals. These five subsets are marked as A, B, C, D, and E. Cross-validation is applied to handle the problem of insufficient data. We set the ratio of the trainng set and validation set to 4:1, as we can train the model for five times. The results of 5-fold cross-validation are listed in Table \ref{validationresult}. From the table, the average accuracy reaches 92.06\% and the other three metrics are 91.81\%, 7.43\%, and 8.49\%, respectively. 

Except for the third group, the accuracies of the other four groups are more than 90\%. In the third group, the accuracy is only 87.50\%, severely lowering the average performance. The other three metrics are also not very good. We investigate the reasons behind it from the corresponding data. We find the EMG signal waves are different sometimes even when they belong to the same motion of one person. Two reasons may lead to this situation. Firstly, the user may not perform the action in the sampling time, therefore time drifting will lead to incorrect labeling, which misleads our system. Secondly, during the data collection step, some participants perform the movements unnaturally, which may produce unqualified data and affect the performance of our system. These reasons also lead to the fluctuations of these metrics in the results of cross validation.

\begin{table}[]
  \caption{Hyperparameters}
  \label{final}
  \begin{tabular}{@{}llll@{}}
  \toprule
  Parameter            & Value & Parameter       & Value               \\ \midrule
  Number of CNN layers & 3     & Filter number   & {[}16, 32, 32{]}    \\
  Learning rate        & 0.002 & Dropout rate    & {[}0.1, 0.2, 0.2{]} \\
  Batch size           & 32    & Number of epochs & 20                  \\ \bottomrule
  \end{tabular}
\end{table}

    
    

\begin{table}[]
  \centering
  \caption{Cross-validation results}
  \label{validationresult}
  \begin{tabular}{@{}cccccc@{}}
  \toprule
  Train set     & Test set    & Accuracy & TAR     & FAR  & FRR   \\ \midrule
  A, B, C, D    & E           & 95.33\%  & 95.83\% & 6.62\% & 4.67\%  \\
  A, C, D, E    & B           & 91.41\%    & 90.22\%   & 8.13\% & 9.78\%  \\
  A, B, D, E    & C           & 87.50\%    & 84.70\%   & 9.11\% & 15.29\% \\
  B, C, D, E    & A           & 92.69\%    & 95.77\%   & 8.12\% & 4.23\%  \\
  A, B, C, E    & D           & 93.37\%    & 92.50\%   & 5.17\% & 7.50\%  \\
  \hline
  \hline
  \multicolumn{2}{c}{Average} & 92.06\%    & 91.81\%   & 7.43\% & 8.29\%  \\ \bottomrule

\end{tabular}
\end{table}

\subsection{Rotation-Independence Verification}
\begin{figure}
  \setlength{\abovecaptionskip}{-0.cm}
  \setlength{\belowcaptionskip}{-0.cm}
  \centering
  \includegraphics[width=0.45\textwidth]{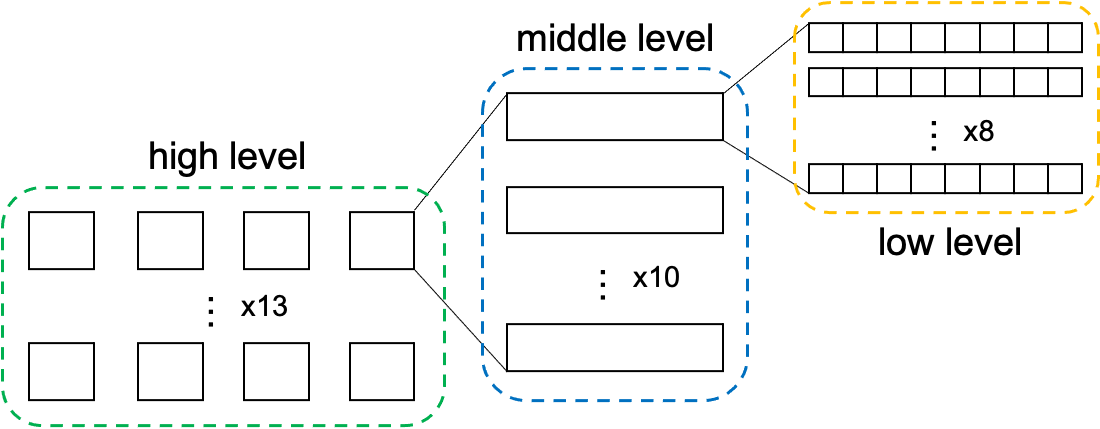}\\
  \caption{Data structure in the rotation-independence verification.}
  \label{datastructure}   
\end{figure}

We conduct several experiments to verify the rotation-independence of EmgAuth. The data structure we use in this section is shown in Figure \ref{datastructure}. In the high level part, the number of rows represents the number of people (13 in our dataset), and the number of columns represents the four wearing positions collected by each person. The middle level part shows each item in the high level has ten pieces of data and the low level part describes the shape of the data which is $10\times8\times400$. The number 8 and 400 are the same meaning with the previous introduction. Therefore, we treat the $8\times400$ matrix as a valid data and there are in total, 280 pieces of data to verify the rotation-independence feature of EmgAuth.

\begin{figure}
  \setlength{\abovecaptionskip}{-0.cm}
  \setlength{\belowcaptionskip}{-0.cm}
  \centering
  \includegraphics[width=0.48\textwidth]{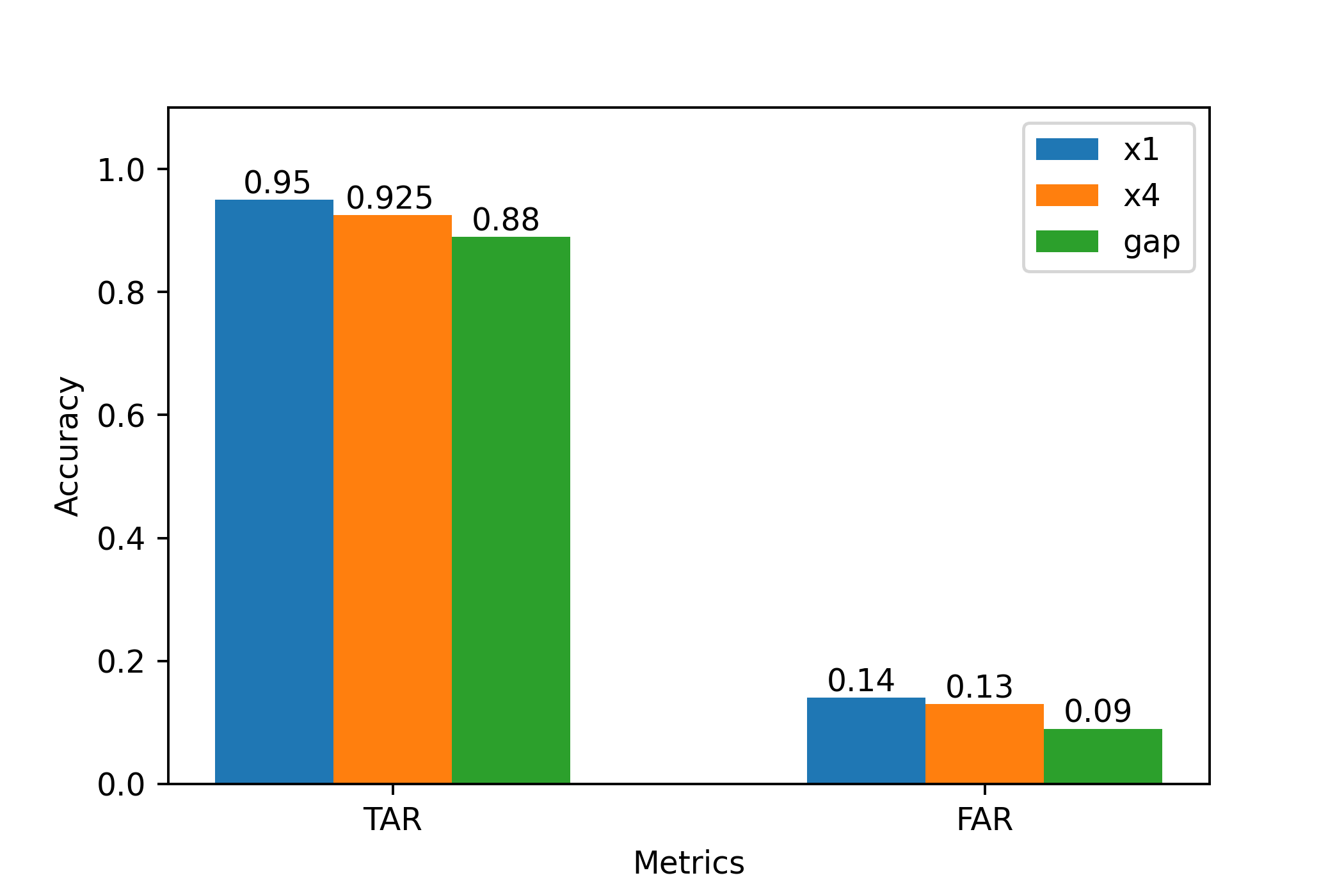}\\
  \caption{TAR and FAR in different positions.}
  \label{rotate}   
\end{figure} 

For the same user, we use the original data to compare with the rotated data with to verify the effectiveness of EmgAuth, when authenticating the same user, in other words, testing the TAR. Specifically, for the same user, we pair the EMG data of one position with the other three positions. In this way, we can obtain 420 pairs of EMG signals. Then we input them into the fine-tuned Siamese network and compare the output with the threshold, which is 0.5. If the output value is less than 0.5, the authentication is considered as successful.

For different users, we use the original data of one user to compare the rotated data of other users to verify whether EmgAuth can correctly identify after rotation, in other words, testing the FAR. Specifically, we pair the EMG data of one position from one user with the four positions from other users, the rest operation is the same with the above. For the result, if the output is bigger the 0.5, we can say the rotation does not mislead EmgAuth to recognize different users as the same one.

Figure \ref{rotate} presents the results of TAR and FAR when the Myo armband rotates at different angles. x1 means the armband rotates one sensor position and x4 means the armband rotates four sensor position. Gap means the armband rotates to a gap position. We can see that TAR is the highest when the armband rotates one sensor position, which is about 0.95. When the rotation angle is 180 degrees, the TAR drops a little to 0.925. EmgAuth performs worst when the armband rotates to a gap position, the corresponding TAR is just 0.88. The reason is that when we apply data augmentation, the expanded data are all in integer multiples of the sensor position and the gap position is not counted. However, the width of gap is much less than sensor's, leading to an acceptable result. Unlike TAR, FAR shows the reverse trend. It reaches the lowest when the position is in the gap. The reason is similar with the above analysis, the model does not see the EMG data in the gap during the training process, which naturally believes two unknown EMG signals are from different users.

\section{Influencing factors in real scenarios}
The reliability of EmgAuth under various working conditions is critical for real-world deployment. In this section, we discuss the scenarios that EmgAuth might encounter in practice, including non-sitting scenarios, different enrollment actions, left-handed suitability. We also investigate the impact of smartphone type and the unlocking speed.

\subsection{Performance when standing}
\begin{figure}[]
  \setlength{\abovecaptionskip}{-0.cm}
  \setlength{\belowcaptionskip}{-0.cm}
  \centering
  \includegraphics[width=0.47\textwidth]{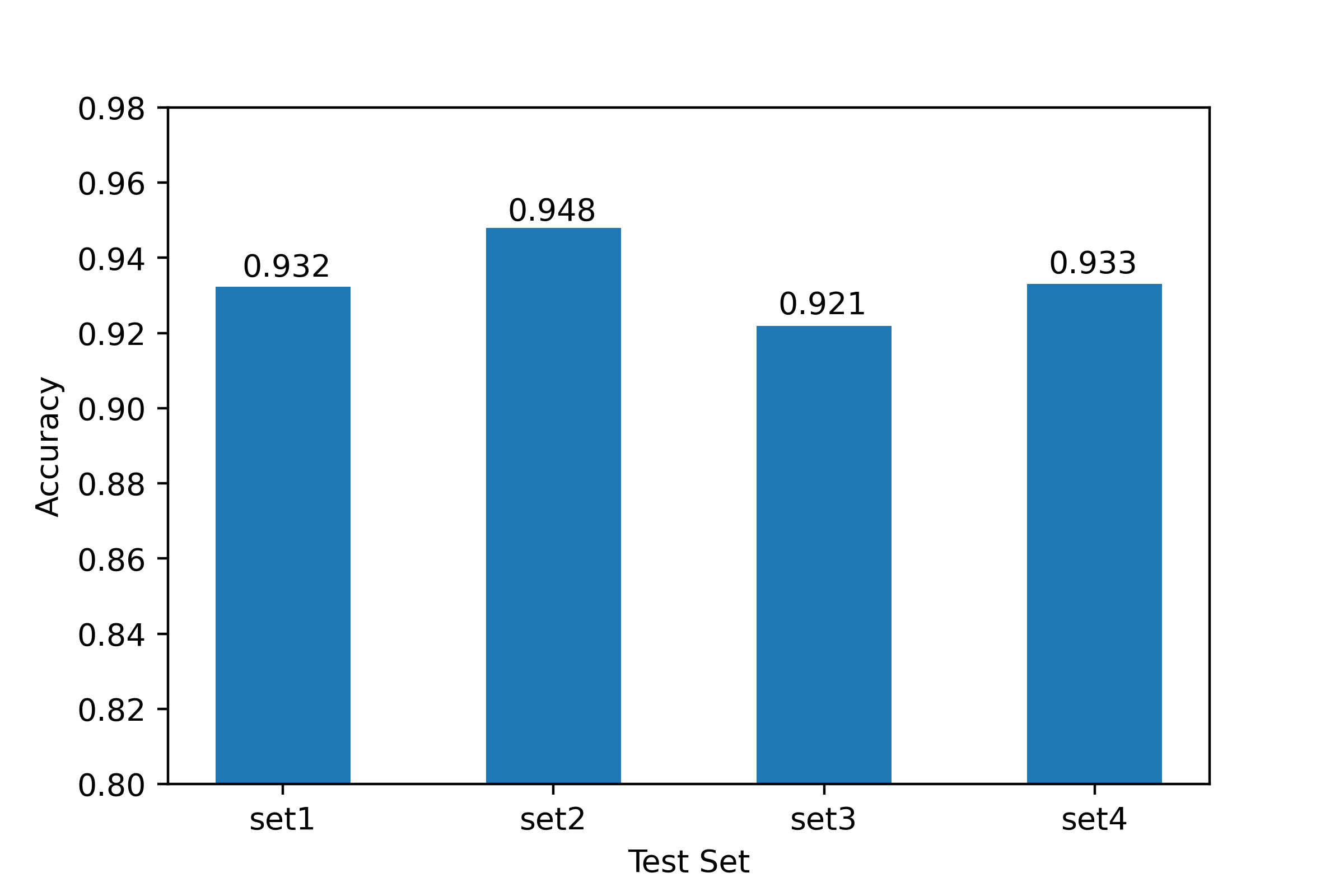}
  \caption{The accuracy of EmgAuth in non-sitting scene. }
  \label{stand}
\end{figure}

In addition to using smartphones when sitting, it is also common to use it while standing. To prove EmgAuth works in this scenario, we use the 130 valid pieces of data (13 participants) that was previously mentioned in the data collection part to train the tiny Siamese network. We choose 9 people for the training set and the other 4 people for testing, and we leverage cross-validation to repeat this experiment 5 times. Considering that the action speed is faster than the sitting scenes, we adjust the sampling time and change the data shape from (8, 400, 1) to (8, 242, 1), the other hyperparameters remain unchanged. The experimental results are shown in Figure \ref{stand}. The average accuracy rate exceeds 92.5\%, which means EmgAuth can deal with the standing scenario. This also means EmgAuth has a good expandability, as long as there is suitable training data.
In consideration of watching a device screen when walking is not safe, we do not test the EmgAuth's performance for this scene.

\subsection{Impact of different enrollment actions}
\begin{figure}
  \setlength{\abovecaptionskip}{-0.cm}
  \setlength{\belowcaptionskip}{-0.cm}
  \centering
  \includegraphics[width=0.3\textwidth]{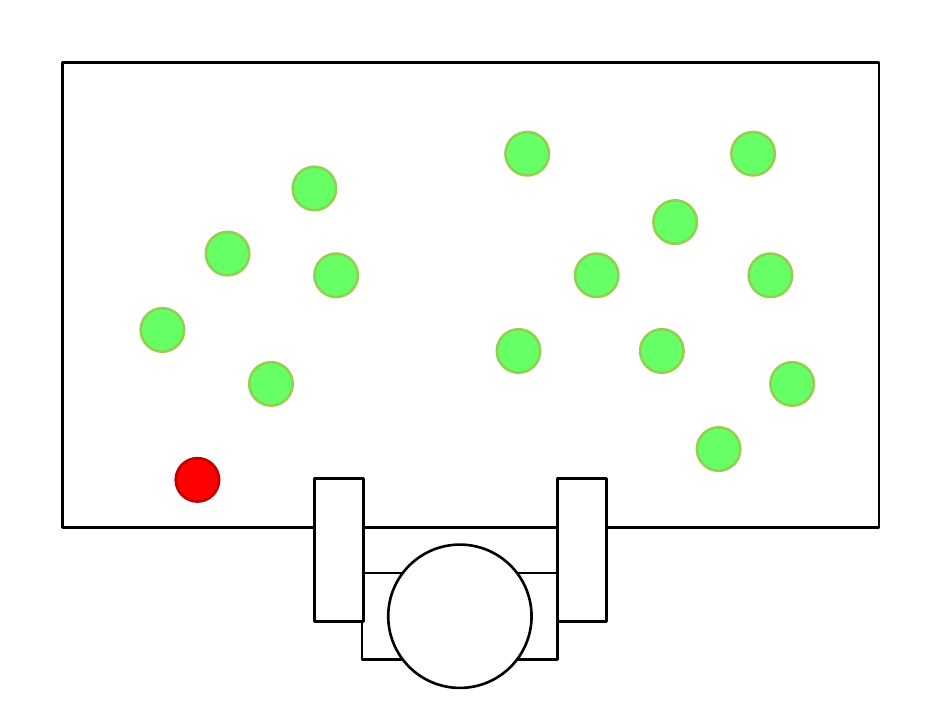}\\
  \caption{Random positions and the authentication results.}
  \label{random}   
\end{figure}

As mentioned before, there are in total four different smartphone positions that we use during the data collection phase. However, in the real scenario, the position of the smartphone is random. In this section, we would like to investigate whether EmgAuth can handle other positions besides the above four locations. We invite five participants to do this experiment. Firstly, they are allowed to place their smartphones as they want (except the above four positions) and we record the positions of them. Then they take the enrollment step to store their EMG signals. The results of the authentication are shown in Figure \ref{random}. The green markers are the positions that EmgAuth authenticates correctly while the red markers are the positions that our system detects by mistake. There is just one position that EmgAuth fails and the accuracy reaches 93.33\%. From this experiment, we can conclude that our deep learning model learns the features well and has an excellent possibility for generalization. We also find that no matter where the smartphone is placed, the finger-level movements are similar when the user grabs his or her smartphone and it is the main reason why the system has good capacity to be generalized.

\subsection{Impact of smartphone size and weight}
To investigate the influence of different types of smartphones, we invite four participants with four different smartphones. We design five sets of trials and every trial corresponds to a specific smartphone. In each trial, the four participants are asked to use the same smartphone to do the enrollment step and authentication step. Then in the next experiment, they change to another smartphone at the same time. We do not set attackers in this experiment so TAR is the only metric to measure the performance. The brands and parameters are listed in Table \ref{smartphone}, as well as the experiment result. 
For the results of TAR, the overall performance is good except for Huawei Honor 10. The TAR of Honor is just 80\% while the others are all more than 90\%. The reason is that the size of Honor 10 is much bigger than the other three smartphones. A participant who gets used to the standard size of a certain smartphone will find it hard to adopt a bigger size one within a short time, which leads to the varying motions in the trials. In addition, we do not find the weight of the smartphone to be an influencing factor of EmgAuth. Therefore, we conclude that EmgAuth is device-independent.

\begin{table}[]
  \centering
  \caption{Different smartphones and the corresponding TAR}
  \label{smartphone}
  \begin{tabular}{@{}llll@{}}
  \toprule
  Product         & Weight (g) & Size (mm) & TAR (\%) \\ \midrule
  iPhone Xs       & 177       & 70.9$\times$143.6$\times$7.7         & 95  \\
  iPhone 7        & 138       & 67.2$\times$138.3$\times$7.1         & 90  \\
  Huawei Honor 10 & 230       & 85$\times$177$\times$7.65         & 80  \\
  Xiaomi 8        & 175       & 74.8$\times$154.9$\times$7.6         & 90  \\ \bottomrule
  \end{tabular}
\end{table}

\subsection{Performance on left hand}
In real life, there are left-handed people as well as those who are right-handed, and this section is to evaluate whether EmgAuth can accommodate with this scenario. We invite two left-handed people to help us perform this experiment. The data collection process is identical with the right hand scenario except for the wearing position. They are required to do the enrollment first and repeat an action five times during the authentication phase. From the experiment, our system fails just once among the ten times. Therefore, we can say that EmgAuth is also suitable for left-handed people. 

\subsection{Unlocking speed}
In this section, we test the speed of EmgAuth on both the server and the Android smartphone. The whole unlocking process includes four steps: 1) loading stored EMG metrics, 2) data segmentation, 3) making pairs, and 4) model calculation. Among the above four steps, loading data consumes the most amount of time because it is an I/O operation. Therefore, we load the EMG matrices in memory after a user finishes the enrollment phase to accelerate the process. Then we test the time both on the server and a Xiaomi 8 smartphone. When the process runs on a server machine, the authentication latency is about 0.048 s. Due to the limited computing resource of the smartphone, the authentication latency of the simulation application running on a Xiaomi 8 smartphone is about 0.16 s, which fulfills the requirement of real-time unlocking.

\section{Discussion}
In this work, we use EMG signals to enable smartphone unlocking without any additional actions, which is a promising supplement to the existing unlocking (or authentication) methods. Unlike other static biometric features, EMG signals are dynamic and changes with body movement, but follows a specific pattern for the same person due to unique and individual muscle structure. Therefore, it naturally enables authentication methods which are hard to acquire without the owner noticicing, thus reduces the risk of privacy leakage. Unlike other research that involves the Myo armband, we are inspired by image augmentation in the field of computer vision and propose EMG signal-based augmentation to free calibration. In other words, people can wear the armband freely rather than fixing to a set position. Moreover, we collect an EMG dataset from 53 people, including sitting and non-sitting scenarios, which could be used for further research.

We notice that not every time users pick up their smartphones to use them, sometimes they just want to put them in their pockets or change their device's positions. In these scenarios, they do not expect their smartphones to be unlocked. On the one hand, the unlocked phone in the pocket may make a call by mistake. On the other hand, if a user moves the phone on the table and leaves it without locking it again, it may lead to privacy leakage. We perform preliminary experimentation for this situation and find that EmgAuth does not trigger unlocking when the user does the above two motions. The reason is that, when we train the Siamese network, the data we labeled starts from picking up the smartphone but is stopped at the point the user is looking at it. In other words, the authentication process is completed by looking at phone, not putting it into the pocket or moving it from one place to another place. If EmgAuth receives wrong EMG signal sequences, the built-in Siamese network will find it and output a big value that is much bigger than the threshold, so the smartphone will not be unlocked by mistake.

EmgAuth allows users to directly use their smartphone devices instead of entering passcodes or waiting for other recognition systems, such as fingerprint or facial recognition systems to finish authenticating, and EmgAuth will not be affected if a user is wearing gloves or masks, which could be used in more environments, especially during Covid-19-like pandemics. 

EmgAuth has to rely on the Myo armband to obtain EMG signals, which is troublesome for real daily life use. People would not like to wear additional devices just to make their smartphones safer. However, we aim to propose a prototype system to verify the feasibility of using EMG signals to unlock smartphones without calibration and pre-designed actions. We believe this paper could give other researchers ideas to explore more biometric signals that can be used for authentication, combining with the powerful feature extraction ability of deep learning to achieve more innovative applications. Also, if the EMG sensor could be integrated into smartwatches one day, EmgAuth will be easy to use and has the potential to unlock other personal digital devices such as laptops, tablets, or even smart home accessories.

\section{Conclusion}
We present EmgAuth, an EMG-based smartphone unlocking system, which leverages EMG signals and a Siamese network to unlock smartphones. In particular, when training the Siamese network, we design a special data augmentation technique to make the system resilient to the rotation of the armband, which lets the system be free of calibration. We conduct an experimentation with 53 participants for collecting a dataset and design a convolutional Siamese network for analysing the EMG signals. Our system can authenticate users effectively with an average TAR of 91.81\% while keeping an average FAR of 7.43\%. Extensive experiments are conducted to test the rotation-independence feature, demonstrate the usability of EmgAuth for smartphones with different sizes and at different locations, as well as users with different postures. To estimate the speed, we measure the latency both on a server and an Android smartphone, the results fulfill the requirement of real-time smartphone unlocking. 

Although the experimental results are promising, the limitation of EmgAuth also exists. First, it is not convenient to wear an armband for unlocking a smartphone. Second, our system may not work well in the humid environment as the EMG signal becomes unstable if the skin surface is wet. In the future, we will evaluate the performance of EmgAuth in a long-term stability study. In particular, we will design a smaller EMG sensor to make the system easier to use and discover more application scenarios. 

\ifCLASSOPTIONcompsoc
  \section*{Acknowledgments}
\else
  \section*{Acknowledgment}
\fi

This work was supported by xx. Thanks to xx gave authors the professional idea about data collection.

\ifCLASSOPTIONcaptionsoff
  \newpage
\fi



\bibliographystyle{IEEEtran}
\bibliography{tmc}
%



%

\begin{IEEEbiography}[{\includegraphics[width=1in,height=1.25in,clip,keepaspectratio]{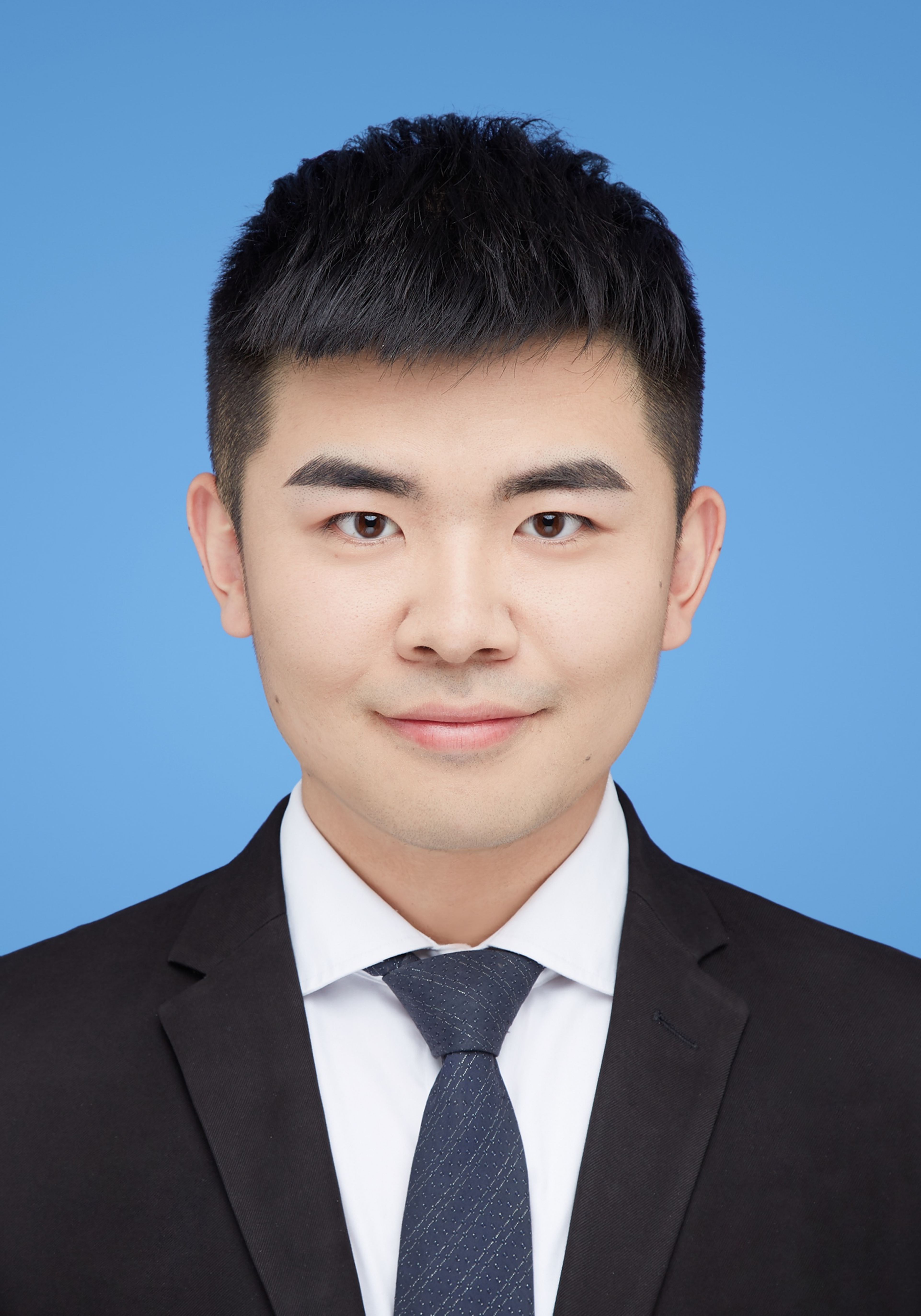}}]{Boyu Fan}
received a BS degree in the Internet of Things Engineering from Taiyuan University of Technology, Taiyuan, China in 2017 and a master's degree from Beihang University, Beijing, China in 2020. He will pursue a PhD degree on Computer Science at the University of Helsinki, Finland. His research interests include Mobile Computing, Internet of Things, and Machine Learning.
\end{IEEEbiography}

\begin{IEEEbiography}[{\includegraphics[width=1in,height=1.25in,clip,keepaspectratio]{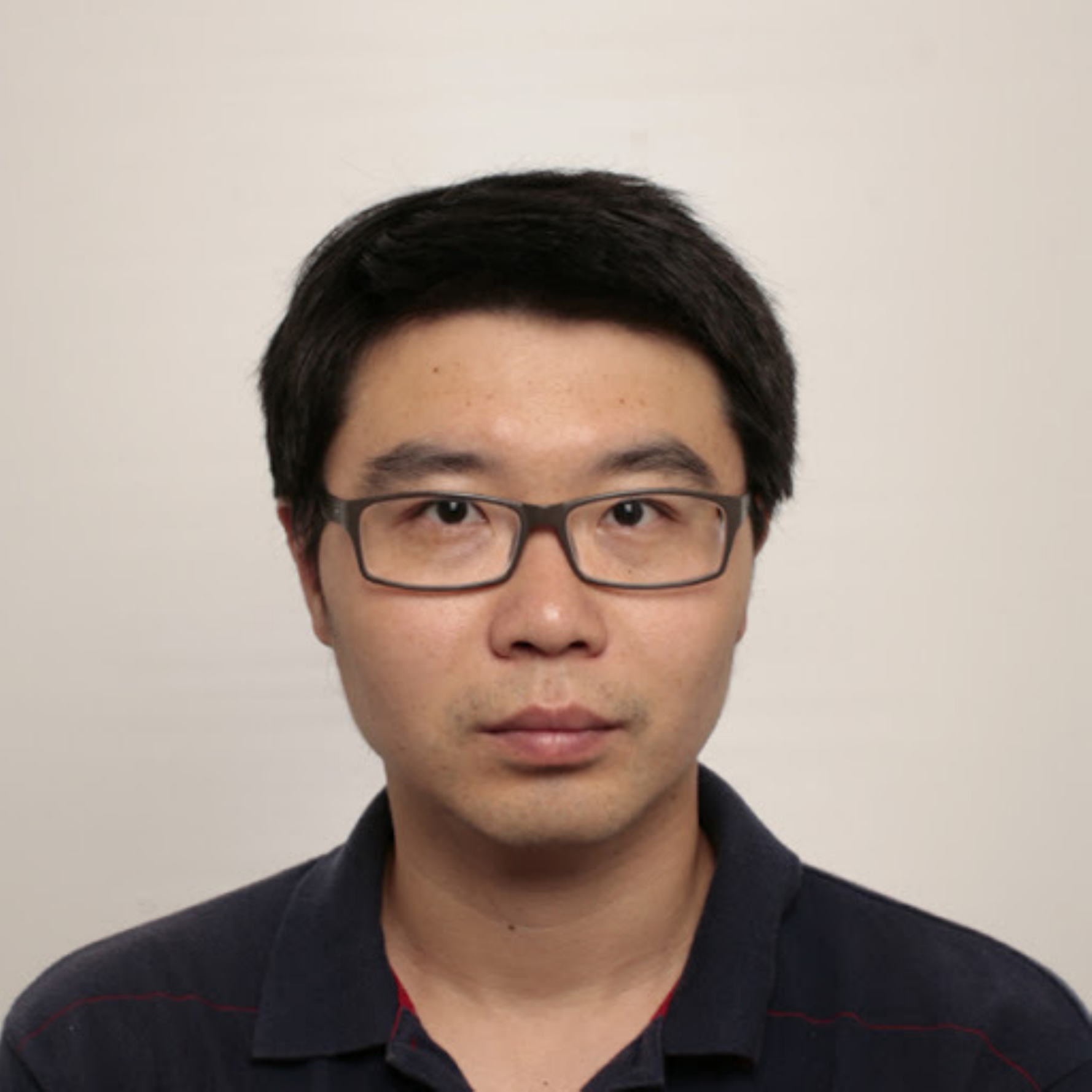}}]{Xiang Su}
  received his Ph.D. in technology from the University of Oulu in 2016. He is currently an Academy of Finland postdoc fellow and a senior postdoctoral researcher in computer science at the University of Helsinki. Dr. Su has extensive expertise on Internet of Things, edge computing, mobile augmented reality, knowledge representations, and context modeling and reasoning. He is a member of IEEE.
\end{IEEEbiography}

\begin{IEEEbiography}[{\includegraphics[width=1in,height=1.25in,clip,keepaspectratio]{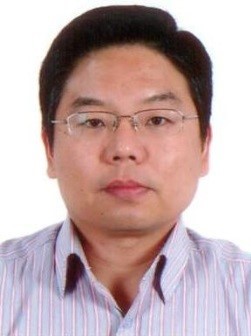}}]{Jianwei Niu}
  received the M.S. and Ph.D. degrees in school of computer science and engineering from Beihang University, Beijing, China, in 1998 and 2002, respectively. He was a visiting scholar at School of Computer Science, Carnegie Mellon University, USA from Jan. 2010 to Feb. 2011. He is a professor in the School of Computer Science and Engineering, BUAA, and an IEEE senior member. His current research interests include mobile and pervasive computing, mobile video analysis and robot operation system.
\end{IEEEbiography}
  
   \begin{IEEEbiography}[{\includegraphics[width=1in,height=1.25in,clip,keepaspectratio]{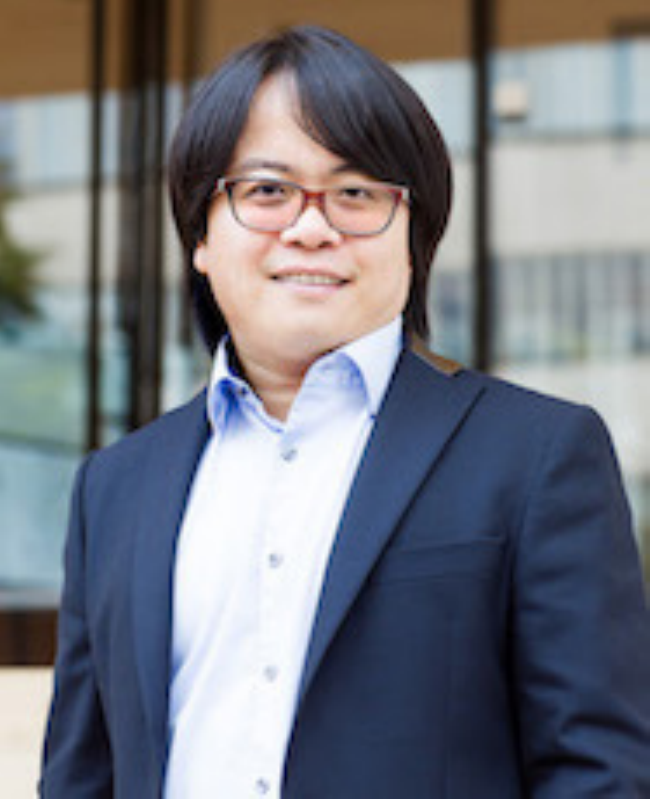}}]{Pan Hui}
   (SM'14-F'18) received his PhD from the Computer Laboratory at the University of Cambridge, and both his Bachelor and MPhil degrees from the University of Hong Kong.
  
   He is the Nokia Chair Professor in Data Science, and Professor of Computer Science at the University of Helsinki. He is also the director of the HKUST-DT System and Media Lab at the Hong Kong University of Science and Technology. He was an adjunct Professor of social computing and networking at Aalto University from 2012 to 2017. He was a senior research scientist and then a Distinguished Scientist for Telekom Innovation Laboratories (T-labs) Germany from 2008 to 2015.  His industrial profile also includes his research at Intel Research Cambridge and Thomson Research Paris from 2004 to 2006. His research has been generously sponsored by Nokia, Deutsche Telekom, Microsoft Research, and China Mobile. He has published more than 300 research papers and with over 17,500 citations. He has 30 granted and filed European and US patents in the areas of augmented reality, data science, and mobile computing. 
  
   He has founded and chaired several IEEE/ACM conferences/workshops, and has served as track chair, senior program committee member, organising committee member, and program committee member of numerous top conferences including ACM WWW, ACM SIGCOMM, ACM Mobisys, ACM MobiCom, ACM CoNext, IEEE Infocom, IEEE ICNP, IEEE ICDCS, IJCAI, AAAI, and ICWSM. He is an associate editor for the leading journals IEEE Transactions on Mobile Computing and IEEE Transactions on Cloud Computing. He is an IEEE Fellow, an ACM Distinguished Scientist, and a member of Academia Europaea.
   \end{IEEEbiography}




\end{document}